\documentclass[10pt,final,journal]{IEEEtran}
\usepackage[switch]{lineno} 
\usepackage{tabularx} 
\usepackage{booktabs}
\usepackage{multirow}
\usepackage{makecell}

\usepackage{stmaryrd}
\usepackage{graphicx}
\usepackage{amsthm} 
\usepackage{amsmath} 
\usepackage{bm}
\usepackage{graphicx}
\usepackage{epstopdf}
\usepackage{color}
\usepackage{float}
\usepackage[colorlinks=false,linkcolor=black,urlcolor=black,bookmarksopen=true, hidelinks]{hyperref}
\usepackage{bookmark}
\usepackage[flushleft]{threeparttable}
\usepackage{stackengine}
\usepackage{scalerel}

\usepackage{xcolor,soul}

\usepackage{cite}

\ifCLASSINFOpdf
\else
\fi
\usepackage{amsfonts,amssymb}

\usepackage{amsmath}
\usepackage{algorithm}
\usepackage{algpseudocode}
\ifCLASSOPTIONcompsoc
 \usepackage[caption=false,font=normalsize,labelfont=sf,textfont=sf]{subfig}
\else
 \usepackage[caption=false,font=footnotesize]{subfig}
\fi
\begin{document}
%
\title{Optimal Localization with Sequential Pseudorange Measurements for Moving Users in a Time Division Broadcast Positioning System}

\author{Sihao~Zhao, \textit{Member, IEEE},
		Xiao-Ping~Zhang, \textit{Fellow, IEEE},
        Xiaowei~Cui,
        and~Mingquan~Lu
\thanks{This work was supported in part by the Natural Sciences and Engineering Research Council of Canada (NSERC), Grant No. RGPIN-2020-04661. \textit{(Corresponding author: Xiao-Ping Zhang.)}}
\thanks{S. Zhao, X.-P. Zhang are with the Department of Electrical, Computer and Biomedical Engineering, Ryerson University, Toronto, ON M5B 2K3, Canada (e-mail: sihao.zhao@ryerson.ca; xzhang@ee.ryerson.ca).}
\thanks{X. Cui is with the Department of Electronic Engineering,
	Tsinghua University, Beijing 100084, China (e-mail: cxw2005@tsinghua.edu.cn).}
\thanks{M. Lu is with the Department of Electronic Engineering,
	Beijing National Research Center for Information Science and Technology, Tsinghua University, Beijing 100084, China. (e-mail: lumq@tsinghua.edu.cn).}
\thanks{Copyright (c) 20xx IEEE. Personal use of this material is permitted. However, permission to use this material for any other purposes must be obtained from the IEEE by sending a request to pubs-permissions@ieee.org.}
}

\markboth{}%
{Shell \MakeLowercase{\textit{et al.}}: Bare Demo of IEEEtran.cls for IEEE Journals}
%




\maketitle

\begin{abstract}
In a time division broadcast positioning system (TDBPS), a user device (UD) determines its position by obtaining sequential time-of-arrival (TOA) or pseudorange measurements from signals broadcast by multiple synchronized base stations (BSs). The existing localization method using sequential pseudorange measurements and a linear clock drift model for the TDPBS, namely LSPM-D, does not compensate the position displacement caused by the UD movement and will result in position error. In this paper, depending on the knowledge of the UD velocity, we develop a set of optimal localization methods for different cases. First, for known UD velocity, we develop the optimal localization method, namely LSPM-KVD, to compensate the movement-caused position error. We show that the LSPM-D is a special case of the LSPM-KVD when the UD is stationary with zero velocity. Second, for the case with unknown UD velocity, we develop a maximum likelihood (ML) method to jointly estimate the UD position and velocity, namely LSPM-UVD. Third, in the case that we have prior distribution information of the UD velocity, we present a maximum \textit{a posteriori} (MAP) estimator for localization, namely LSPM-PVD. We derive the Cram\'er-Rao lower bound (CRLB) for all three estimators and analyze their localization error performance. We show that the position error of the LSPM-KVD increases as the assumed known velocity deviates from the true value. As expected, the LSPM-KVD has the smallest position error while the LSPM-PVD and the LSPM-UVD are more robust when the prior knowledge of the UD velocity is limited. Numerical results verify the theoretical analysis on the optimality and the positioning accuracy of the proposed methods.

\end{abstract}



%
\IEEEpeerreviewmaketitle

\section{Introduction}\label{Introduction}
%
%
%
%
\IEEEPARstart{L}{o}calization utilizing a variety of measurements including time-of-arrival (TOA), direction-of-arrival (DOA), received signal strength (RSS), etc., extracted from wireless signals has drawn more and more attention in research and industrial fields \cite{guvenc2009survey,luo2019novel,liu2007survey,wang2012novel,shao2014efficient,xue2019locate}. In a wireless broadcast positioning system, the base stations (BSs) at known places transmit signals and a user device (UD) or receiver needs to receive such signals to obtain TOA or pseudorange measurements for position determination \cite{yassin2016recent,fallahi2011robust,kaplan2005understanding}. One of the advantages of such a broadcast positioning system is that an unlimited amount of users are supported because UDs passively receive signals. Furthermore, the users in such a system are safer because they are silent and thus difficult to be detected.


Among the broadcast positioning schemes, code division and frequency division schemes are widely adopted to enable simultaneous signal transmission from multiple BSs. Global Positioning System (GPS) and Glonass are two example positioning systems who employ the code division and frequency division schemes, respectively \cite{hofmann2007gnss,misra2006global}. However, the code division scheme suffers from the near-far effect  especially for a small region such as indoor cases where a user may be very close to a BS \cite{hui1998successive,picois2014near}, and the frequency division scheme requires very stable carrier frequency and occupies several frequency bands \cite{myung2006single}.

The time division (TD) scheme has no near-far effect compared with code division and no requirement on frequency filtering compared with frequency division. With the advent of new measurement techniques such as ultra-wide band (UWB) and acoustic sensors, TD scheme is becoming more widely studied and more pervasively adopted in both academic and industrial fields \cite{de2014ultrasonic,leugner2016comparison,beuchat2019enabling,hamer2018self,tiemann2019scalability,shi2019range}.


In a TD broadcast positioning system (TDBPS), BSs with known positions transmit signals in separate time slots to the air and UDs receive these signals to locate themselves \cite{hamer2018self,tiemann2019scalability}. Different from TDBPS, in code or frequency division system, a UD receive multiple signals from different BSs at the same time and extract the concurrent pseudorange measurements for positioning \cite{jiang2015locata,khalife2018navigation,khyam2018design,an2020distributed}. A plenty of research works on localization methods based on concurrent pseudorange or TOA measurements are studied, which lay a foundation for the localization method for the TDBPS. The positioning method presented in \cite{kaplan2005understanding,tsui2005fundamentals,foy1976position} is a widely adopted approach that utilizes the extracted pseudorange measurements concurrently from four or more satellites or BSs by the UD. Various forms of such method are extensively discussed in literature, such as weighting the measurements to reduce errors in multi-path or other noisy situations \cite{wang2017robust,kim2010distance}, and squaring the measurements to simplify the method to a closed form \cite{bancroft1985algebraic,cheung2004least,chan1994simple}. However, all these conventional positioning methods assume that multiple measurements at the same reception time are available to determine the UD position. With this assumption, only the UD position and clock offset at the reception time are unknowns to be solved, while the user’s dynamics including its velocity and clock drift are not estimated and compensated. In contrast to the systems with concurrent pseudorange measurements available, the UD in the TDBPS receives signal from only one BS and obtains a single pseudorange measurement at one time. During multiple measurement receptions, a moving UD changes its position, and the relative clock offset between the UD and BSs also changes due to the clock drift or oscillator frequency difference. Therefore, the above-mentioned conventional localization method based on concurrent pseudorange measurements (LCPM) is not suitable for applications in the TDBPS.

Filtering methods such as Kalman filter and particle filter have the capability to work with sequential TOA measurements and can bridge gaps caused by insufficient measurements and track a moving UD by incorporating historical measurements and model on the UD dynamics \cite{1292138,jwo2007adaptive,jia2016target,cano2019kalman,hu2017quantized,cai2019asynchronous}. However, these filtering methods require the UD motion model throughout the entire tracking period. If the actual UD movement deviates from the preset model, there will be extra estimation errors. They also suffer from convergence speed, inaccurate state noises and measurement outliers. They are not designed specifically for the TDBPS.

In \cite{segura2012ultra,segura2010experimental}, a TDBPS with code division scheme to distinguish different BSs is presented and the conventional LCPM algorithm is directly adopted. This LCPM method can give acceptable positioning results when the measurement time slot is very short and the UD is moving slowly. However, if the time length for localization and/or the UD velocity increases, large error or even divergence will appear in the positioning result due to large UD displacement and/or accumulated clock offset. The work in \cite{saad2012high} assumes the movement of a UD is negligible and uses historical TOA measurements from the same BS to linearly estimate the clock offset, so that the conventional LCPM algorithm can be directly used. This method cannot deal with moving UDs with large displacement during the localization period and lack of previous measurements will lead to inaccurate clock offset estimation and thus degrade the accuracy of the positioning result. A TD signal transmission scheme is proposed by \cite{pelka2016s}, in which the LCPM method is used for localization with beforehand measured clock drift. However, measuring the clock drift beforehand is time-consuming for real-world applications, and the actual drift may deviate from the measured value due to the instability of the oscillator, leading to extra position error.

A localization method utilizing sequential pseudorange measurements with an extra state of clock drift, namely LSPM-D hereinafter, is adopted in recent research works. In \cite{ledergerber2015robot,hamer2018self}, a TD positioning scheme is presented and the LSPM-D method is used to support multiple robots for localization. They treat the multiple measurements as sequential ones and add the clock drift term as an extra state. Shi et al. propose a TDBPS that completes joint localization and synchronization based on UWB hardware \cite{shi2019blas}. They also adopt an extra clock drift state to compensate the accumulating clock offset during the measurement period. However, the above literature simply ignore the user velocity during multiple-measurement reception. This condition is not practical in the real world especially for a moving UD and/or a long time length of multiple-measurement reception in the TDBPS. The unmodeled user movement will cause extra position errors, which will degrade the localization performance, especially in the case of large UD speed or displacement during the measurement period.

In this paper, we formulate the UD localization problem in the TDBPS with a linear motion model, which has constant velocity during a short period. For the case with known UD velocity, we propose the optimal localization method, namely LSPM-KVD, to compensate the user displacement during the measurement period. We show that it is a generalized form of the conventional LSPM-D, and fixes the problem of extra position error due to lacking velocity compensation. An iterative LSPM-KVD (ILSPM-KVD) algorithm is presented to solve the localization problem. In order to handle the case with unknown UD velocity, we propose a joint position and velocity estimator, namely LSPM-UVD, and design an iterative localization algorithm, namely ILSPM-UVD. We then propose a maximum \textit{a posteriori} (MAP) estimator-based localization method utilizing prior knowledge on the UD velocity, namely LSPM-PVD, which is shown to be a generalized form of the LSPM-KVD and LSPM-UVD. We conduct error analysis and derive the Cram\'er-Rao lower bound (CRLB) for the three methods. Position error analysis shows that the localization accuracy of the LSPM-KVD is better than that of the LSPM-D, the positioning accuracy of the LSPM-KVD degrades when the known velocity deviates from the true value, and the localization accuracy of the LSPM-PVD lies in the range from the lower bound that equals the CRLB of the LSPM-KVD to the upper bound given by the LSPM-UVD. Simulation results show the optimality and verify the theoretical analysis for all the three methods.

The rest of the paper is organized as follows. In Section II, the model of TDBPS is presented and the localization problem based on pseudorange measurement with linear UD motion is formulated. Three optimal localization methods for the cases with different prior knowledge on the UD velocity, namely LSPM-KVD, LSPM-UVD and LSPM-PVD, and their iterative algorithms are proposed in detail in Section III. The positioning performances of the three proposed methods are analyzed in Section IV. Simulation results to evaluate the performances for the proposed LSPM-KVD, LSPM-UVD and LSPM-PVD are given in Section V. Finally, Section VI draws the conclusion of this paper.
 
Main notations are summarized in Table \ref{table_notation}.

\begin{table}[!t]
	\caption{Notation List}
	\label{table_notation}
	\centering
	\begin{tabular}{l p{5.5cm}}
		\toprule
		lowercase $x$&  scalar\\
		bold lowercase $\boldsymbol{x}$ & vector\\
		bold uppercase $\bm{X}$ & matrix\\
		$\hat{x}$, $\hat{\boldsymbol{x}}$ & estimate of a variable\\
		$\Vert \boldsymbol{x} \Vert$ & Euclidean norm of a vector\\
		$\Vert \boldsymbol{x}\Vert _{\bm{W}}^2$ & square of Mahalanobis norm, i.e., $\boldsymbol{x}^T\bm{W}\boldsymbol{x}$\\
		$\mathrm{tr}(\bm{X})$ & trace of a matrix\\
		$\left|\bm{X}\right|$ & determinant of a matrix\\
		$[\bm{X}]_{u,:}$, $[\bm{X}]_{:,v}$ &the $u$-th row and the $v$-th column of a matrix, respectively\\
		$[\bm{X}]_{u:v,m:n}$ &sub-matrix with the $u$-th to the $v$-th rows and the $m$-th to the $n$-th columns\\
		$[\bm{X}]_{u,v}$ &entry at the $u$-th row and the $v$-th column of a matrix\\
		$[\boldsymbol{x}]_{u}$ &the $u$-th element of a vector\\
		$\mathbb{E}[\cdot]$ & expectation operator \\
		$\mathrm{diag}(\cdot)$ & diagonal matrix with the elements inside\\
		$M$ & number of pseudorange measurements or observed BSs\\
		$N$ & dimension of all the position and velocity vectors, i.e., $N=2$ in 2D case and $N=3$ in 3D case\\
		$i$, $j$ & index of the measurements (or the observed BS)\\
		$\bm{O}_{M \times N}$ & $M \times N$ matrix with all-zero entries\\
		$\bm{I}_{N}$ &$N \times N$ identity matrix\\
		$\boldsymbol{q}_{i}$ & position vector of the $i$-th observed BS\\
		$\boldsymbol{p}$ &  unknown position vector of UD\\
		$\boldsymbol{v}$ &  velocity vector of UD\\
		$\rho_{i}$ & the $i$-th pseudorange measurement between UD and BS\\
		$b$, $d$ &  clock offset and clock drift between UD and BS\\
		$t_L$ &  time instant for localization\\
		$\boldsymbol{\theta}$ &  parameter vector\\
		$\varepsilon$, $\sigma^2$ &  pseudorange measurement noise and variance\\
		$\mathcal{F}$ &  Fisher information matrix\\
		$\bm{W}$ & weighting matrix\\
		$\bm{\Sigma}$ & variance matrix\\
		$\bm{G}$ & design matrix\\
		$\mu$& position bias\\
		$\bm{Q}$ & position error variance matrix\\		
		\bottomrule
	\end{tabular}
\end{table}

\section{Problem Statement} \label{problem}
\subsection{TDBPS Model}
In a TDBPS, a UD only receives the signals transmitted from BSs to determine its own position parameters. The locations of BSs are known. We denote the position of the $i$-th observed BS by $\boldsymbol{q}_i$, where $i=1,\cdots,M$, and $M$ is the total number of the observed BSs whose signals are received by the UD. All BSs are synchronized and their time is considered as the reference for the entire system and is named \textit{system time} and denoted by $t$ hereinafter. The UD states include the user position denoted by $\boldsymbol{p}$, and the clock offset denoted by $b$. The dimension of all the position vectors is $N$ (e.g., $N=2$ in 2D case and $N=3$ in 3D case), i.e., $\boldsymbol{q}_i, \boldsymbol{p}\in \mathbb{R}^N$.

\begin{figure}
	\centering
	\includegraphics[width=1\linewidth]{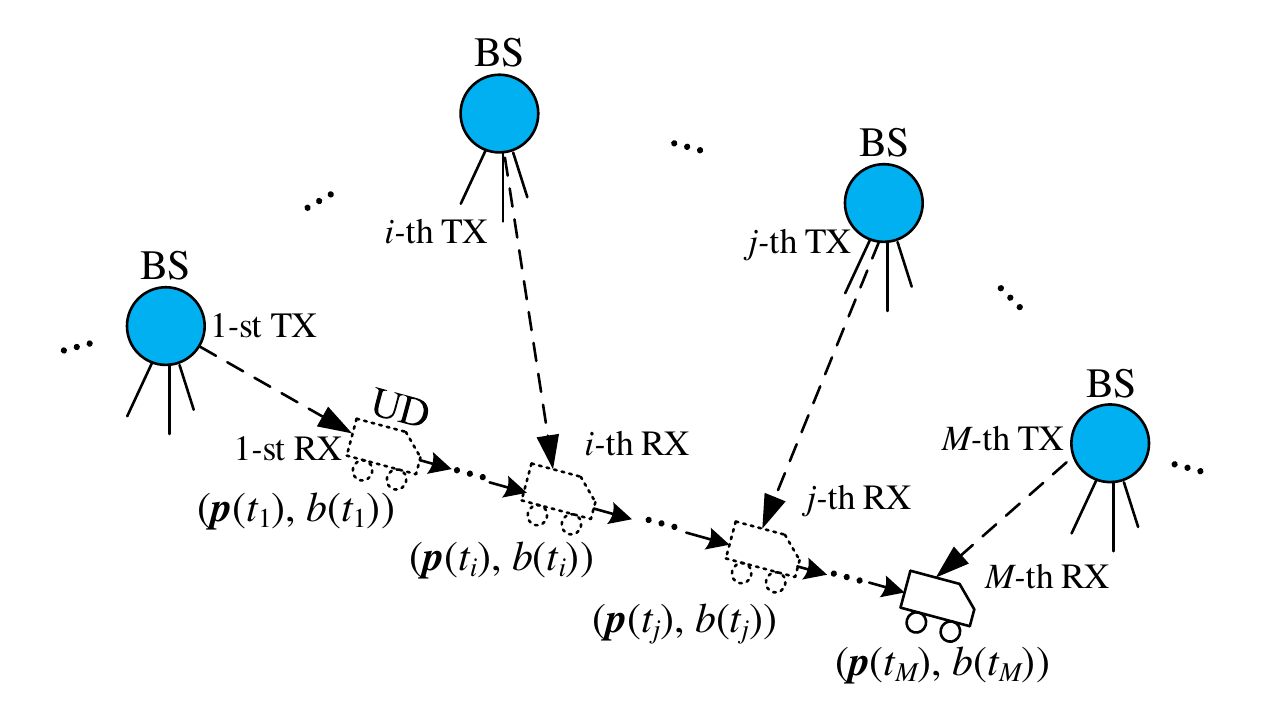}
	\caption{Sequential signal transmission and reception in a TDBPS. The position and clock offset of a moving UD change with time. The UD obtains $M$ sequential measurements to complete localization.
	}
	\label{fig:rxstate}
\end{figure}

Fig. \ref{fig:rxstate} shows that BSs broadcast and a UD receives signals in a sequential manner while the UD position and clock offset are changing with time. Upon signal reception, the UD measures the TOA based on its local clock that is not synchronized with the system time. When the UD moves, its position changes with time. Thus, the UD position is a function of $t$, and the position at the $i$-th reception instant is denoted by $\boldsymbol{p}(t_i)$. Note that two clock sources with identical nominal oscillation frequency generate sine waves with a slightly different frequencies when working in the real world. It causes a time-varying relative clock offset between two devices \cite{zucca2005clock}. Therefore, the UD clock offset with respect to the system time also changes with time and thus is denoted by $b(t_i)$ at the $i$-th reception instant.


\subsection{TDBPS Pseudorange-based Localization Problem with Short-time Linear Motion Model}
The broadcast signal is stamped with the BS transmission time. Once received, the TOA measured by the UD local clock is obtained. Due to asynchronization between the UD and BS, only a pseudorange between them can be obtained, i.e., there is an extra term caused by the UD clock offset with respect to the BSs. The pseudorange measurement is given by 
\begin{equation} \label{eq:pri}
\rho_i(t_i) =\Vert \boldsymbol{q}_i - \boldsymbol{p}(t_i) \Vert + c\cdot b(t_i) + \varepsilon_i \text{,}\; i=1,\cdots,M\text{,}
\end{equation}
where $\rho_i$ is the pseudorange measurement between the UD and the $i$-th observed BS, $t_i$ is the reception time given by the system clock, $c$ is the signal propagation speed, and $\varepsilon_i$ is the measurement noise, which is independent between different BSs and follows a zero mean Gaussian distribution with a variance of $\sigma_i^2$, i.e., $\varepsilon_i \sim \mathcal{N}(0,\sigma_i^2)$. We note that signals from different BSs may undergo different propagation paths, and therefore the noise variance may be different between measurements.


Similar to (\ref{eq:pri}), the pseudorange measurement with respect to the $j$-th observed BS is written as
\begin{align} \label{eq:prj}
\rho_j(t_j) =& \left\Vert \boldsymbol{q}_j - \boldsymbol{p}(t_j) \right\Vert + c\cdot b(t_j) + \varepsilon_j \nonumber\\
=&\left\Vert \boldsymbol{q}_j - \left(\boldsymbol{p}(t_i)+\int_{t_i}^{t_j}\boldsymbol{v}(t) dt\right) \right\Vert \nonumber\\
&+ c\cdot \left(b(t_i)+\int_{t_i}^{t_j}d(t) dt\right) + \varepsilon_j
\text{,} \; i,j=1,\cdots,M\text{,}
\end{align}
where $\boldsymbol{v}(t)$ is the UD velocity and $d(t)$ is the clock drift between the UD and BSs that varies with time due to the frequency deviation of the UD oscillator.

During a short time interval and/or with a small user velocity, the UD position can be linearly modeled as the sum of the initial position and the product of a constant velocity and the time period. We select a time instant $t_L$ at which the UD position is computed and denote the time difference between the current instant of measurement and the localization time by $\Delta t_j=t_j-t_L$. Then, the UD position $\boldsymbol{p}$ at time instant $t_j$ given by (\ref{eq:prj}) becomes
\begin{equation} \label{eq:posvel}
\boldsymbol{p}(t_j) = \boldsymbol{p}(t_L) + \boldsymbol{v}(t_L) \cdot \Delta t_j \text{.}
\end{equation}

With respect to the UD clock drift, it is also treated as constant during a short interval. Hence, the clock offset and the drift terms in (\ref{eq:prj}) are given by the linear model as
\begin{equation} \label{eq:clockb}
b(t_j) = b(t_L) + d(t_L) \cdot \Delta t_j \text{.}
\end{equation}

Based on the known position of the $j$-th BS denoted by $\boldsymbol{q}_j$, the clock model given in (\ref{eq:clockb}), and the UD's short-time movement model in (\ref{eq:posvel}), the pseudorange measurement at a certain instant $t_j$ given by (\ref{eq:prj}) can be linearly connected to the parameters at instant $t_L$, which is written as
\begin{align} \label{eq:rhoandpos}
\rho_j(t_j) = &\left\Vert \boldsymbol{q}_j - \left(\boldsymbol{p}(t_L) + \boldsymbol{v}(t_L) \cdot \Delta t_j\right)\right\Vert \nonumber\\
&+ c\cdot b(t_L) + c\cdot d(t_L) \cdot \Delta t_j + \varepsilon_j
\text{.}
\end{align}

Note that in (\ref{eq:rhoandpos}), the BS position term $\boldsymbol{q}_j$ is known. By using the linear model given by (\ref{eq:posvel}) and (\ref{eq:clockb}), we only have to estimate $\boldsymbol{p}$, $\boldsymbol{v}$, $b$ and $d$ at the selected time instant $t_L$ rather than at every $t_i$ and $t_j$. Thus, the required number of measurements and computational complexity are both reduced.

The localization problem for the TDBPS is to estimate the UD position $\boldsymbol{p}$ at a selected time instant $t_L$ using $M$ pseudorange measurements given by (\ref{eq:pri}) and (\ref{eq:rhoandpos}). For simplicity, the conventional LSPM-D method ignores the velocity term. By doing this, the problem is simplified at the cost that the final position error will grow with an increasing UD speed due to lack of displacement compensation. In order to tackle this problem, we will propose a set of optimal localization methods in this paper.


\section{Proposed Optimal Localization Methods for TDBPS}

\subsection{Optimal Localization with Known UD Velocity}
\subsubsection{ML Estimator for Localization}\label{usermodel}
When the UD velocity during the measurement period is known by some means such as an external sensor, the parameters to be estimated in (\ref{eq:pri}) and (\ref{eq:rhoandpos}) are the UD position and the clock offset. We can use a maximum likelihood (ML) estimator to solve this localization problem. We name this method by localization based on sequential pseudorange measurement with a known UD velocity and a clock drift state to be estimated - LSPM-KVD.

The parameters to be estimated in this case is denoted by $\boldsymbol{\theta}_{K}$, which contains the UD position $\boldsymbol{p}$, clock offset $b$ and drift $d$, i.e., 
$$
\boldsymbol{\theta}_{K}=\left[\boldsymbol{p}^T,b,d\right]^T \text{,}
$$
and $\boldsymbol{\theta}_{K} \in \mathbb{R}^{N+2}$. Here, we include the signal propagation speed in the clock offset and drift terms for simple illustration. The subscript ``$K$'' represents the LSPM-KVD method. The parameters to be estimated are all at time instant $t_L$ in the following text, and thus the time index for the corresponding variables is omitted in the following text without loss of clarity.

The relationship between the pseudorange measurements and the unknown parameters in a collective form reads
\begin{equation} \label{eq:MLvectorpr}
\boldsymbol{\rho} = h(\boldsymbol{\theta}_{K}) + \boldsymbol{\varepsilon} \text{,}
\end{equation}
where $\boldsymbol{\rho} \in \mathbb{R}^M$ is a vector containing all pseudorange measurements, i.e., $\boldsymbol{\rho}=[\rho_1,\cdots,\rho_M]^T$,  and $\boldsymbol{\varepsilon}$ is the vector containing all the pseudorange measurement noises, i.e., $\boldsymbol{\varepsilon}=[\varepsilon_1,\cdots,\varepsilon_M]^T$, and $h(\cdot)$ is a nonlinear mapping function as given by (\ref{eq:rhoandpos}) for a single pseudorange measurement, i.e.,
$$\left[h(\boldsymbol{\theta}_K)\right]_i=\Vert \boldsymbol{q}_i - (\boldsymbol{p} +\boldsymbol{v} \Delta t_i) \Vert + b +d \Delta t_i \text{,}\; i=1,\cdots,M \text{.}
$$

Recall that the measurement noises of different BSs are independent and follow a Gaussian distribution, the parameters can then be estimated by solving the weighted least squares (WLS) minimization problem as
\begin{equation} \label{eq:MLminimizer}
\hat{\boldsymbol{\theta}}_{K}=\text{arg}\min\limits_{{\boldsymbol{\theta}}_{K}} \left\Vert\boldsymbol{\rho} - \mathit{h}({\boldsymbol{\theta}}_{K})\right\Vert_{\bm{W}_{\rho}}^2
\text{,}
\end{equation}
where $\hat{\boldsymbol{\theta}}_{K}$ is the estimator, and  $\bm{W}_{\rho}$ is a positive-definite diagonal weighting matrix given by
\begin{equation} \label{eq:weightmatrho}
	\bm{W}_{\rho}= \bm{\Sigma}^{-1}_{\rho}= \mathrm{diag}\left(\frac{1}{\sigma_1^2},\cdots,\frac{1}{\sigma_M^2}\right) \text{,}
\end{equation}
with $\bm{\Sigma}_{\rho}=\mathrm{diag}\left({\sigma_1^2},\cdots,{\sigma_M^2}\right)$ being the pseudorange measurement noise variance matrix, and $\Vert \boldsymbol{x}\Vert _{\bm{W}}^2=\boldsymbol{x}^T\bm{W}\boldsymbol{x}$. We know that ML estimator is asymptotically unbiased \cite{kay1993fundamentals}, and thus, $\hat{\boldsymbol{\theta}}_{K}$ converges to the true value in probability.

At this stage, the localization problem in such a TDBPS reduces to determining the parameter $\boldsymbol{\theta}_{K}$ based on the non-linear equation given by (\ref{eq:MLvectorpr}). This estimator using sequential pseudorange measurements $\boldsymbol{\rho}$ and the known UD velocity is named as LSPM-KVD.

We note that when the UD velocity is zero, the nonlinear function $h(\cdot)$ in (\ref{eq:MLvectorpr}) becomes that of the conventional LSPM-D. Thus, the LSPM-D can be considered as a special case of the LSPM-KVD when the UD is stationary. This will be further shown in the error analysis in Section \ref{comparison}.

\subsubsection{Iterative LSPM-KVD Algorithm}
We adopt the commonly used Gauss-Newton method \cite{kaplan2005understanding,misra2006global} to construct the iterative algorithm for the LSPM-KVD as well as the other two proposed methods. First, to linearize the equation set given by (\ref{eq:MLvectorpr}), we conduct a Taylor series expansion at the estimate point of
$
\check{\boldsymbol{\theta}}_{K}=\left[\check{\boldsymbol{p}}^T,\check{b},\check{d}\right]^T \text{,}
$
where $\check{\boldsymbol{p}}^T$, $\check b$, and $\check d$ are estimates for ${\boldsymbol{p}}^T$, $b$, and $d$ at time $t_L$, respectively.

When the estimate is close to the true parameter, the high order terms are small enough to be ignored. We retain the first order term of the Taylor expansion, and (\ref{eq:MLvectorpr}) becomes
\begin{equation} \label{eq:KVDtaylor}
	\boldsymbol{\rho} = \mathit{h}(\check{\boldsymbol{\theta}}_{K}) + \left(\frac{\partial \mathit{h}(\boldsymbol{\theta}_{K})}{\partial \boldsymbol{\theta}_{K}}|_{\boldsymbol{\theta}_{K}=\check{\boldsymbol{\theta}}_{K}}\right)\left(\boldsymbol{\theta}_{K}-\check{\boldsymbol{\theta}}_{K}\right)+\boldsymbol\varepsilon \text{.}
\end{equation}

After ignoring the higher order terms in (\ref{eq:KVDtaylor}), we obtain a linear equation set. We define the error vector:
$$\Delta\boldsymbol{\theta}_{K} \triangleq \boldsymbol{\theta}_{K}-\check{\boldsymbol{\theta}}_{K} \text{,}
$$
and the residual vector:
\begin{equation}\label{eq:KVDresidual}
\boldsymbol{r}_K \triangleq \boldsymbol{\rho} - \mathit{h}(\check{\boldsymbol{\theta}}_{K})= \check{\bm{G}}_K \cdot \Delta\boldsymbol{\theta}_{K}+\boldsymbol\varepsilon \text{,}
\end{equation}
where $\check{\bm{G}}_K$ is the estimation of the design matrix ${\bm{G}}_K$, which is determined by the true value of the parameter $\boldsymbol{\theta}_K$, and
\begin{align} \label{eq:KVDGderivative}
[\check{\bm{G}}_K]_{i,:}=\left[\frac{\partial \mathit{h}(\boldsymbol{\theta}_K)}{\partial \boldsymbol{\theta}_K}|_{\boldsymbol{\theta}_{K}=\check{\boldsymbol{\theta}}_{K}}\right]_{i,:} =\left[-\check{\boldsymbol{e}}_{Ki}^T,1,\Delta t_i \right] \text{,}
\end{align}
\begin{equation} \label{eq:KVDLOS}
	\check{\boldsymbol{e}}_{Ki}=\frac{\boldsymbol{q}_i - \check{\boldsymbol{p}}-{\boldsymbol{v}} \Delta t_i}{\Vert \boldsymbol{q}_i - \check{\boldsymbol{p}}-{\boldsymbol{v}} \Delta t_i\Vert } \text{,}
\end{equation}
with  $[\cdot]_{i,:}$ denoting the $i$-th row of a matrix, and $\check{\boldsymbol{e}}_K$ representing the estimated unit line-of-sight (LOS) vector from the UD to the BS.

The design matrix ${\bm{G}}_K$ is defined as
\begin{align} \label{eq:trueKVDGderivative}
	[{\bm{G}}_K]_{i,:}
	\triangleq\left[-{\boldsymbol{e}}_{i}^T,1,\Delta t_i \right] \text{,}
\end{align}
where $\boldsymbol{e}$ is the unit LOS vector given by
\begin{equation} \label{eq:trueKVDLOS}
	{\boldsymbol{e}_i}=\frac{\boldsymbol{q}_i - {\boldsymbol{p}}-{\boldsymbol{v}} \Delta t_i}{\Vert \boldsymbol{q}_i - {\boldsymbol{p}}-{\boldsymbol{v}} \Delta t_i\Vert } \text{.}
\end{equation}

The WLS estimate of the error vector $\Delta\boldsymbol{\theta}_K$,  denoted by $\Delta \check{\boldsymbol{\theta}}_K$, is given by
\begin{equation} \label{eq:KVDleastsquare}
\Delta\check{\boldsymbol{\theta}}_K=(\check{\bm{G}}_K^T\bm{W}_{\rho}\check{\bm{G}}_K)^{-1}\check{\bm{G}}_K^T\bm{W}_{\rho}\boldsymbol{r}_K \text{.}
\end{equation}

Then, the estimated parameter vector can be updated by
\begin{equation} \label{eq:KVDtheta}
\check{\boldsymbol{\theta}}_K \leftarrow \check{\boldsymbol{\theta}}_K + \Delta \check{\boldsymbol{\theta}}_K \text{.}
\end{equation}

The updated parameter estimation given by (\ref{eq:KVDtheta}) is then used to calculate the value of the function $\mathit{h}(\cdot)$. After that, the residual vector $\boldsymbol{r}_K$, design matrix $\bm{G}_K$ and error vector $\Delta\boldsymbol{\theta}_K$ are updated iteratively. Once the estimated error vector $\Delta\check{\boldsymbol{\theta}}_K$ is sufficiently small, the iteration converges. The Gauss-Newton iterative LSPM-KVD (ILSPM-KVD) for the TDBPS is summarized in Algorithm 1.

Note that the estimate parameter obtained from the iterative algorithm is biased due to the non-linearity of the problem \cite{teunissen1990nonlinear,wang2000integrating}. However, when the UD is far from the BS and the measurement noise is small, this non-linearity-caused bias is small enough to be ignored, as analyzed in \cite{yan2008feasibility}. Therefore, in this work, we can safely use the proposed algorithm to solve the localization problem in the TDBPS, as shown in the numerical simulations in Section \ref{simulation}.

\begin{algorithm}
	\caption{ILSPM-KVD}
	\begin{algorithmic}[1]
		\State Input: pseudorange measurement $\boldsymbol{\rho}$ and noise variance $\Sigma_{\rho}$, time instant of localization $t_L$, observed BSs' positions $\boldsymbol{q}_i$, $i=1,\cdots,M$, known UD velocity $\boldsymbol{v}$, initial parameter estimate $\check{\boldsymbol{\theta}}_{K0}=[\check{\boldsymbol{p}}_{0}^T, \check b_0, \check d_0]^T$, maximum iterative time $iter$, and convergence threshold $thr$.
		\For {$s=1:iter$}
		\State Calculate unit LOS vector $\check{\boldsymbol{e}}_{Ki}$ based on (\ref{eq:KVDLOS}), $i=1,\cdots,M$
		\State Compute residual vector $\boldsymbol{r}_K$ using (\ref{eq:KVDresidual})
		\State Form design matrix $\check{\bm{G}}_K$ based on (\ref{eq:KVDGderivative})
		\State Calculate estimated error vector $\Delta \check{\boldsymbol{\theta}}_K$ using (\ref{eq:KVDleastsquare})
		\State Update parameter estimate $\check{\boldsymbol{\theta}}_{Ks} = \check{\boldsymbol{\theta}}_{K(s-1)} + \Delta \check{\boldsymbol{\theta}}_K$
		\If {$\Vert \Delta \check{\boldsymbol{\theta}}_K \Vert<thr$}
		\State Exit \textbf{for} loop
		\EndIf
		\EndFor
		\State Output: $\check{\boldsymbol{\theta}}_{Ks}$
	\end{algorithmic}
\end{algorithm}

\subsection{Optimal Localization with Unknown UD Velocity}
\subsubsection{Joint Position and Velocity Estimator}
We discuss how to localize a UD when its velocity is known in the previous sub-section. For the case with unknown UD velocity, we propose the optimal localization method that jointly estimate the position and velocity, namely LSPM-UVD.

The parameter vector to be estimated is denoted by $\boldsymbol{\theta}_U$, containing the UD position, clock offset and drift, and velocity, i.e.,
$$
\boldsymbol{\theta}_U=\left[\boldsymbol{p}^T,b,d,\boldsymbol{v}^T\right]^T \text{,}
$$
where the subscript ``$U$'' represents the LSPM-UVD method and will be used in other variables.

Similar to (\ref{eq:MLvectorpr}), the relationship between the pseudorange measurements and the unknown parameters is given by
\begin{equation} \label{eq:UVDvectorpr}
\boldsymbol{\rho} = \mathit{h}(\boldsymbol{\theta}_U) + \boldsymbol{\varepsilon} \text{.}
\end{equation}
where the function $h(\cdot)$ has the same form as that of (\ref{eq:MLvectorpr}), but becomes a nonlinear $\mathbb{R}^{2N+2}\rightarrow\mathbb{R}^{M}$ mapping function due to the $N$-dimension unknown velocity. 

The parameters can be obtained by solving the minimization problem as
\begin{equation} \label{eq:UVDMLminimizer}
\hat{\boldsymbol{\theta}}_{U}=\text{arg}\min\limits_{{\boldsymbol{\theta}}_{U}}\left\Vert\boldsymbol{\rho} - \mathit{h}({\boldsymbol{\theta}}_{U})\right\Vert_{\bm{W}_{\rho}}^2
\text{,}
\end{equation}
where $\hat{\boldsymbol{\theta}}_{U}$ is the estimator.


\subsubsection{Iterative LSPM-UVD Algorithm} \label{ILSPM-UVD}
The process of the iterative LSPM-UVD (ILSPM-UVD) algorithm is similar to that of the ILSPM-KVD as given by Algorithm 1 using Gauss-Newton method. The main differences include the LOS vector, design matrix and the estimated error vector.

We denote the design matrix by ${\bm{G}}_U$ as given by
\begin{align} \label{eq:trueUVDGderivative}
	[{\bm{G}}_U]_{i,:}=\left[-{\boldsymbol{e}}_i^T,1,\Delta t_i, -{\boldsymbol{e}}_i^T \Delta t_i\right] \text{.}
\end{align}


The estimated unit LOS vector of the LSPM-UVD, denoted by $\check{\boldsymbol{e}}_{Ui}$, is
\begin{equation} \label{eq:UVDLOS}
\check{\boldsymbol{e}}_{Ui}=\frac{\boldsymbol{q}_i - \check{\boldsymbol{p}}-\check{\boldsymbol{v}} \Delta t_i}{\Vert \boldsymbol{q}_i - \check{\boldsymbol{p}}-\check{\boldsymbol{v}} \Delta t_i\Vert } \text{,}
\end{equation}
where $\check{\boldsymbol{v}}$ is the estimated UD velocity, which is different from the true velocity in $\check{\boldsymbol{e}}_{K}$ given by (\ref{eq:KVDLOS}).

The estimated design matrix denoted by $\check{\bm{G}}_U$ is
\begin{align} \label{eq:UVDGderivative}
[\check{\bm{G}}_U]_{i,:}=\left[-\check{\boldsymbol{e}}_{Ui}^T,1,\Delta t_i, -\check{\boldsymbol{e}}_{Ui}^T \Delta t_i\right] \text{.}
\end{align}

The estimated error vector $\Delta\check{\boldsymbol{\theta}}_U$  is given by
\begin{equation} \label{eq:UVDleastsquare}
\Delta\check{\boldsymbol{\theta}}_U=(\check{\bm{G}}_U^T\bm{W}_{\rho}\check{\bm{G}}_U)^{-1}{\check{\bm{G}}_U^T}\bm{W}_{\rho}\boldsymbol{r}_U \text{.}
\end{equation}

The process of the proposed ILSPM-KVD is summarized in Algorithm 2.

\begin{algorithm}
	\caption{ILSPM-UVD}
	\begin{algorithmic}[1]
		\State Input: pseudorange measurement $\boldsymbol{\rho}$ and noise variance $\Sigma_{\rho}$, time instant of localization $t_L$, observed BSs' positions $\boldsymbol{q}_i$, $i=1,\cdots,M$, initial parameter estimate $\check{\boldsymbol{\theta}}_{U0}=[\check{\boldsymbol{p}}_{0}^T, \check b_0, \check d_0, \check{\boldsymbol{v}}_0^T]^T$, maximum iterative time $iter$, and convergence threshold $thr$.
		\For {$s=1:iter$}
		\State Calculate unit LOS vector $\check{\boldsymbol{e}}_{Ui}$ based on (\ref{eq:UVDLOS}), $i=1,\cdots,M$
		\State Compute residual vector $\boldsymbol{r}_U = \boldsymbol{\rho} - \mathit{h}(\check{\boldsymbol{\theta}}_{U})$
		\State Form design matrix $\check{\bm{G}}_U$ based on (\ref{eq:UVDGderivative})
		\State Calculate estimated error vector $\Delta\check{\boldsymbol{\theta}}_U$ using (\ref{eq:UVDleastsquare})
		\State Update parameter estimate $\check{\boldsymbol{\theta}}_{Us} = \check{\boldsymbol{\theta}}_{U(s-1)} + \Delta \check{\boldsymbol{\theta}}_U$
		\If {$\Vert \Delta \check{\boldsymbol{\theta}}_U \Vert<thr$}
		\State Exit \textbf{for} loop
		\EndIf
		\EndFor
		\State Output: $\check{\boldsymbol{\theta}}_{Us}$
	\end{algorithmic}
\end{algorithm}

\subsection{Optimal Localization with Prior Distribution of UD Velocity}
\subsubsection{MAP Estimator for Localization}
The localization cases with known and unknown UD velocity are investigated in the previous two sub-sections. In practice, the UD can be equipped with some low-cost sensors, which may not be accurate enough but can still provide some information on the velocity. Thus, we are able to obtain some prior knowledge on the UD velocity, e.g., its distribution, which is beneficial to improve the positioning accuracy if properly incorporated. In this sub-section, we propose a generalized MAP estimator with prior distribution on velocity, namely LSPM-PVD, to handle this case and include the LSPM-KVD and LSPM-UVD as its special cases. We use a Gaussian distributed velocity as an example to design the localization algorithm.

The parameters to be estimated, denoted by $\boldsymbol{\theta}_P$, contain the position, velocity, clock offset and drift, identical with that of the LSPM-UVD, i.e.,
$$
\boldsymbol{\theta}_P=\boldsymbol{\theta}_U=\left[\boldsymbol{p}^T,b,d,\boldsymbol{v}^T\right]^T \text{.}
$$
where the subscript ``$P$'' represents the LSPM-PVD method and will be applied to other related variables.

The prior distribution function of the UD velocity is denoted by $g(\boldsymbol{v})$. Then, we have the MAP estimator as
\begin{equation} \label{eq:MAPestimator}
\hat{\boldsymbol{\theta}}_P=\text{arg}\max\limits_{{\boldsymbol{\theta}}_P} f(\boldsymbol{\rho}|{\boldsymbol{\theta}}_P)g(\boldsymbol{v})  \text{,}
\end{equation}
where $\hat{\boldsymbol{\theta}}_P$ is the estimator, and the conditional probability density function of $\boldsymbol{\rho}$ given $\boldsymbol{\theta}_P$ is 
\begin{align}\label{eq:VDPDF}
f(\boldsymbol{\rho}|\boldsymbol{\theta}_P)=
\frac{\exp\left(-\frac{1}{2} \Vert\boldsymbol{\rho} - \mathit{h}({\boldsymbol{\theta}_P})\Vert_{\boldsymbol\Sigma_{\rho}^{-1}}^2\right)}{(2\pi)^{\frac{M}{2}} |\bm{\Sigma}_{\rho}|^{\frac{1}{2}}}.
\end{align}

We can see from (\ref{eq:MAPestimator}) that when we know the UD velocity, the prior distribution $g(\boldsymbol{v})$ becomes a delta function. In this case, (\ref{eq:MAPestimator}) becomes
$
\hat{\boldsymbol{\theta}}_K=\text{arg}\max\limits_{{\boldsymbol{\theta}}_K} f(\boldsymbol{\rho}|{\boldsymbol{\theta}}_K)  \text{,}
$
which is equivalent to (\ref{eq:MLminimizer}). Therefore, the LSPM-KVD method becomes a special case of the LSPM-PVD.

In the case without any knowledge on the UD velocity, the prior distribution $g(\boldsymbol{v})$ becomes a constant function, and (\ref{eq:MAPestimator}) degenerates  to $\hat{\boldsymbol{\theta}}_U=\text{arg}\max\limits_{{\boldsymbol{\theta}}_U} f(\boldsymbol{\rho}|{\boldsymbol{\theta}}_U)$, which is equivalent to (\ref{eq:UVDMLminimizer}). This equivalence indicates that the LSPM-UVD is also a special case of the LSPM-PVD.

\subsubsection{Iterative LSPM-PVD Algorithm with Gaussian Distributed UD Velocity}


We assume the UD velocity follows a Gaussian distribution and denote the mean of the velocity by $\overline{\boldsymbol{v}}$, and the variance by $\bm{\Sigma}_v$. The distribution function $g(\boldsymbol{v})$ thereby reads
\begin{equation} \label{eq:VelPDF}
g(\boldsymbol{v}) = \frac{\exp\left(-\frac{1}{2}\left\Vert\boldsymbol{v}-\overline{\boldsymbol{v}}\right\Vert_{\boldsymbol\Sigma_{v}^{-1}}^2\right)}{(2\pi)^{\frac{N}{2}} |\bm{\Sigma}_{v}|^{\frac{1}{2}}}  \text{.}
\end{equation}

By substituting (\ref{eq:VDPDF}) and (\ref{eq:VelPDF}) into (\ref{eq:MAPestimator}), the MAP estimator becomes
\begin{equation} \label{eq:MAPGaussian}
\hat{\boldsymbol{\theta}}_P=\text{arg}\max\limits_{{\boldsymbol{\theta}}_P} \frac{\exp\left(-\frac{1}{2} \Vert\boldsymbol{\rho} - \mathit{h}({\boldsymbol{\theta}}_P)\Vert_{\boldsymbol\Sigma_{\rho}^{-1}}^2-\frac{1}{2}\left\Vert\boldsymbol{v}-\overline{\boldsymbol{v}}\right\Vert_{\boldsymbol\Sigma_{v}^{-1}}^2\right)}{(2\pi)^{\frac{M+N}{2}} |\bm{\Sigma}_{\rho}|^{\frac{1}{2}}|\bm{\Sigma}_{v}|^{\frac{1}{2}}}  \text{.}
\end{equation}

The MAP estimator is equivalent to the WLS minimization problem
\begin{equation} \label{eq:minimizer}
\hat{\boldsymbol{\theta}}_P=\text{arg}\min\limits_{{\boldsymbol{\theta}}_P} \Vert\mathit{\eta}({\boldsymbol{\theta}}_P)\Vert_{\bm{W}}^2 \text{,}
\end{equation}
where
$$
[\eta(\boldsymbol{\theta}_P)]_{1:M}=\boldsymbol{\rho} - \mathit{h}({\boldsymbol{\theta}_P}) \text{,}
$$
$$
[\eta(\boldsymbol{\theta}_P)]_{M+1:M+N}=\overline{\boldsymbol{v}}-\boldsymbol{v} \text{,}
$$
and $\bm{W}$ is the weighting matrix given by
\begin{equation} \label{eq:weightmat}
\bm{W}= \bm{\Sigma}^{-1}=\left[
\begin{matrix}
\bm{\Sigma}_{\rho}^{-1} & \bm{O}_{M\times N}\\
\bm{O}_{N\times M}& \bm{\Sigma}_v^{-1}
\end{matrix}
\right] \text{,}
\end{equation}
with $\bm{\Sigma}$ representing the collection of all the variances.

The iterative algorithm for the LSPM-PVD, namely ILSPM-PVD, is similar to the ILSPM-UVD. However, there are some differences including the estimated design matrix, LOS vector and error vector.

The design matrix for LSPM-PVD is denoted by $\bm{G}_P$. It is given by
\begin{align} \label{eq:trueGderivative}
	&[\bm{G}_P]_{i,:}=\left[-\boldsymbol{e}_{i}^T,1,\Delta t_i, -\boldsymbol{e}_{i}^T\Delta t_i\right] \text{,}\; 1\leq i\leq M \text{,} \nonumber\\
	&[\bm{G}_P]_{M+1:M+N,:}=[\bm{O}_{N\times (N+2)}, \bm{I}_{N}]
	\text{.}
\end{align}

The estimated design matrix $\check{\bm{G}}_P$ is given by
\begin{align} \label{eq:Gderivative}
&[\check{\bm{G}}_P]_{i,:}=\left[-\check{\boldsymbol{e}}_{Pi}^T,1,\Delta t_i, -\check{\boldsymbol{e}}_{Pi}^T\Delta t_i\right] \text{,}\; 1\leq i\leq M \text{,} \nonumber\\
&[\check{\bm{G}}_P]_{M+1:M+N,:}=[\bm{G}_P]_{M+1:M+N,:}
\text{,}
\end{align}
where the estimated LOS vector is equal to that of the LSPM-UVD, i.e.,
\begin{equation} \label{eq:PVDLOS}
\check{\boldsymbol{e}}_{Pi}=\check{\boldsymbol{e}}_{Ui}=\frac{\boldsymbol{q}_i - \check{\boldsymbol{p}}-\check{\boldsymbol{v}} \Delta t_i}{\Vert \boldsymbol{q}_i - \check{\boldsymbol{p}}-\check{\boldsymbol{v}} \Delta t_i\Vert }  \text{.}
\end{equation}
The estimated error vector is given by
\begin{equation} \label{eq:PVDleastquares}
\Delta\check{\boldsymbol{\theta}}_P=(\check{\bm{G}}_P^T\bm{W}_{\rho}\check{\bm{G}}_P)^{-1}\check{\bm{G}}_P^T\bm{W}_{\rho}\boldsymbol{r}_P \text{.}
\end{equation}
The iterative ILSPM-PVD procedure is given by Algorithm 3.

\begin{algorithm}
	\caption{ILSPM-PVD with Gaussian distributed UD velocity}
	\begin{algorithmic}[1]
		\State Input: pseudorange measurement $\boldsymbol{\rho}$ and noise variance $\Sigma_{\rho}$, time instant of localization $t_L$, observed BSs' positions $\boldsymbol{q}_i$, $i=1,\cdots,M$, prior known velocity mean $\overline{\boldsymbol{v}}$ and variance $\bm{\Sigma}_v$, initial parameter estimate $\check{\boldsymbol{\theta}}_{P0}=[\check{\boldsymbol{p}}_{0}^T, \check b_0, \check d_0, \check{\boldsymbol{v}}_0^T]^T$, maximum iterative time $iter$, and convergence threshold $thr$.
		\For {$s=1:iter$}
		\State Calculate unit LOS vector $\check{\boldsymbol{e}}_{Pi}$ based on (\ref{eq:PVDLOS}), $i=1,\cdots,M$
		\State Compute residual vector $\boldsymbol{r}_P = \boldsymbol{\rho} - \mathit{h}(\check{\boldsymbol{\theta}}_{P})$
		\State Form design matrix $\check{\bm{G}}_P$ based on (\ref{eq:Gderivative})
		\State Calculate estimated error vector $\Delta\check{\boldsymbol{\theta}}_P$ using (\ref{eq:PVDleastquares})
		\State Update parameter estimate $\check{\boldsymbol{\theta}}_{Ps} = \check{\boldsymbol{\theta}}_{P(s-1)} + \Delta \check{\boldsymbol{\theta}}_P$
		\If {$\Vert \Delta \check{\boldsymbol{\theta}}_P \Vert<thr$}
		\State Exit \textbf{for} loop
		\EndIf
		\EndFor
		\State Output: $\check{\boldsymbol{\theta}}_{Ps}$
	\end{algorithmic}
\end{algorithm}

\section{Localization Performance Analysis} \label{locanalysis}

\subsection{LSPM-KVD Position Error Analysis} \label{ErrorKVD}
\subsubsection{Position Error Performance} \label{errorKVD}
The position bias, variance and root mean square error (RMSE) of the LSPM-KVD are analyzed. We denote the position error of the LSPM-KVD by $\Delta\boldsymbol{p}_K$, and the bias of the LSPM-KVD, denoted by $\mu_K$, is
\begin{align}
\boldsymbol{\mu}_K=\mathbb{E}\left[\Delta\boldsymbol{p}_K\right]
=\left[\mathbb{E}\left[(\bm{G}_K^T\bm{W}_{\rho}\bm{G}_K)^{-1}\bm{G}_K^T\bm{W}_{\rho}\boldsymbol{\varepsilon}\right]\right]_{1:N}=\boldsymbol{0} \text{.} 
\end{align}



The position error variance and the RMSE are thereby
\begin{align}
\bm{Q}_K&=\mathbb{E}\left[\left(\Delta\boldsymbol{p}_K-\mathbb{E}[\Delta\boldsymbol{p}_K]\right)\left(\Delta\boldsymbol{p}_K-\mathbb{E}[\Delta\boldsymbol{p}_K]\right)^T\right] \nonumber \\
&=\left[(\bm{G}_K^T\bm{W}_{\rho}\bm{G}_K)^{-1}\right]_{1:N,1:N}\text{,} \nonumber \\
RMSE_K &=\sqrt{\Vert\boldsymbol{\mu}_K\Vert^2+\mathrm{tr}(\bm{Q}_K)}=\sqrt{\mathrm{tr}(\bm{Q}_K)} \text{,}
\end{align}
where $\bm{Q}_K$ is the position variance and $RMSE_K$ is the position RMSE.

%

\subsubsection{CRLB Derivation}
We derive the CRLB of the proposed LSPM-KVD in this subsection. When the UD collects $M$ measurements from signals transmitted from BSs, the likelihood function $f(\boldsymbol{\rho}|\boldsymbol{\theta}_K)$ is
\begin{equation} \label{eq:KVDlikelihood}
f(\boldsymbol{\rho}|\boldsymbol{\theta}_K) = \frac{\exp\left(-\frac{1}{2} \Vert\boldsymbol{\rho} - \mathit{h}({\boldsymbol{\theta}_K})\Vert_{\boldsymbol\Sigma_{\rho}^{-1}}^2\right)}{(2\pi)^{\frac{M}{2}} |\bm{\Sigma}_{\rho}|^{\frac{1}{2}}} \text{.}
\end{equation}


The second-order derivative is given by
\begin{align} \label{eq:KVD2ndlikelihood}
&\frac{\partial^2 \ln f(\boldsymbol{\rho}|\boldsymbol{\theta}_K)}{\partial \boldsymbol{\theta}_K \partial \boldsymbol{\theta}_K^T} =\nonumber\\
&-\left(\frac{\partial h(\boldsymbol{\theta}_K)}{\partial \boldsymbol{\theta}_K}\right)^T\bm{\Sigma}_{\rho}^{-1}\frac{\partial h(\boldsymbol{\theta}_K)}{\partial \boldsymbol{\theta}_K}-\left(\boldsymbol{\rho}-h(\boldsymbol{\theta}_K)\right)^T\bm{\Sigma}_{\rho}^{-1}\frac{\partial^2 h(\boldsymbol{\theta}_K)}{\partial \boldsymbol{\theta}_K\partial \boldsymbol{\theta}_K^T} \text{.}
\end{align}

The entries of the Fisher information matrix (FIM) denoted by $\mathcal{F}_K$ are 
\begin{equation} \label{eq:KVDFisherExpectation}
[\mathcal{F}_K]_{i,j}=-\mathbb{E}\left[\frac{\partial^2 \ln f(\boldsymbol{\rho}|\boldsymbol{\theta}_K)}{\partial[\boldsymbol{\theta}_K]_i \partial[\boldsymbol{\theta}_K]_j}\right] \text{,}
\end{equation}
in which $\mathbb{E}[\cdot]$ is the expectation operator, and $[\cdot]_{i,j}$ is the entry of a matrix at the $i$-th row and the $j$-th column.

We take expectation on (\ref{eq:KVD2ndlikelihood}). The FIM is thereby
\begin{equation}\label{eq:KVDFIMvsG}
\mathcal{F}_K = \bm{G}_K^T\bm{W}_{\rho}\bm{G}_K \text{,}
\end{equation}
where $\bm{G}_{K}$ is given by (\ref{eq:trueKVDGderivative}) and $\bm{W}_{\rho}$ is given by (\ref{eq:weightmatrho}).


The CRLB relating to the $i$-th element in the parameter vector $\boldsymbol{\theta}_K$  is then obtained by 
\begin{equation} \label{eq:KVDCRLBFisher}
\mathsf{CRLB}_K([\boldsymbol{\theta}_K]_i)=[\mathcal{F}_K^{-1}]_{i,i} \text{,}
\end{equation}
where  $[\cdot]_{i}$ represents the $i$-th element of a vector.

Eq.~(\ref{eq:KVDCRLBFisher}) gives the CRLB for the proposed LSPM-KVD. The position related CRLB, i.e., the top-left $N$ ($N$=2 for 2D cases and $N$=3 for 3D cases) diagonal entries of the matrix, are of the most interest in the localization problem.

\subsubsection{Position Error with Deviated Assumed Known Velocity} \label{deviatedV}
In practice, the assumed known UD velocity may not be accurate enough, i.e., it may deviate from the true value. We investigate the impact of the velocity deviation on the final position error.

We denote the assumed known UD velocity in such a case by $\tilde{\boldsymbol{v}}$. We use ``$\sim$'' over a symbol to represent the case for deviated assumed UD velocity. The deviation from the true velocity is denoted by $\Delta \boldsymbol{v}=\tilde{\boldsymbol{v}}-\boldsymbol{v}$. Based on (\ref{eq:KVDtaylor}), the deviated velocity-caused error vector, denoted by $\tilde{\boldsymbol{r}}_K$, is given by
\begin{align}\label{eq:resdV}
[\tilde{\boldsymbol{r}}_{K}]_i =\left\Vert \boldsymbol{q}_i - \left(\boldsymbol{p} + \boldsymbol{v}  \Delta t_i\right)\right\Vert  -\left\Vert \boldsymbol{q}_i - \left({\boldsymbol{p}} + \tilde{\boldsymbol{v}}  \Delta t_i\right)\right\Vert
\text{.}
\end{align}

Then, the position bias denoted by $\tilde{\boldsymbol{\mu}}_{K}$ is 
\begin{equation}\label{eq:dPvsdV}
\tilde{\boldsymbol{\mu}}_{K}=\left[(\tilde{\bm{G}}_K^T\bm{W}_{\rho}\tilde{\bm{G}}_K)^{-1}\tilde{\bm{G}}_K^T\bm{W}_{\rho}\tilde{\boldsymbol{r}}_K\right]_{1:N} \text{,}
\end{equation}
where the design matrix $\tilde{\bm{G}}_K$ is given by
\begin{align} \label{eq:KVDGdV}
[\tilde{\bm{G}}_K]_{i,:}=\left[-{\boldsymbol{e}}_{i}^T,1,\Delta t_i \right] \text{.}
\end{align}

The position variance denoted by $\tilde{\bm{Q}}_K$ and the RMSE denoted by ${\stackon[-8pt]{{$RMSE$}}{\vstretch{1.5}{\hstretch{6.2}{\widetilde{\phantom{\;}}}}}}_K$ are given by
\begin{equation} \label{eq:posvardV}
\tilde{\bm{Q}}_K=\left[(\tilde{\bm{G}}_K^T\bm{W}_{\rho}\tilde{\bm{G}}_K)^{-1}\right]_{1:N,1:N} \text{,}
\end{equation}
and
\begin{equation} \label{eq:RMSEdV}
{\stackon[-8pt]{{$RMSE$}}{\vstretch{1.5}{\hstretch{6.2}{\widetilde{\phantom{\;}}}}}}_K  =\sqrt{\Vert\tilde{\boldsymbol{\mu}}_K\Vert^2+\mathrm{tr}(\tilde{\bm{Q}}_K)} \text{,}
\end{equation}
respectively.

\hypertarget{R1}{\textbf{Remark 1}}: The LSPM-KVD is an unbiased estimator with accurately known velocity. However, the position bias of the LSPM-KVD increases approximately linearly with greater speed deviation $\Vert\Delta \boldsymbol{v}\Vert$ from the true UD velocity. The derivation of the position bias in this case is given in Appendix \ref{Appendix1}.

\subsubsection{Comparison with Conventional LSPM-D} \label{comparison}
We compare the position error of the proposed LSPM-KVD with that of the LSPM-D.


Different from the LSPM-KVD, the conventional LSPM-D method simply ignores the UD movement during the measurement period, resulting in a position bias caused by the inaccurate motion model. We apply the subscript ``$D$'' to the variables for the LSPM-D. The position bias denoted by $\boldsymbol{\mu}_{D}$ for the LSPM-D is given by

\begin{equation}\label{eq:pbiasD}
\boldsymbol{\mu}_{D}=\left[(\bm{G}_K^T\bm{W}_{\rho}\bm{G}_K)^{-1}\bm{G}_K^T\bm{W}_{\rho}\boldsymbol{r}_{D}\right]_{1:N} \text{,}
\end{equation}
where the residual vector $\boldsymbol{r}_{D}$ is expressed as
\begin{align} \label{eq:measurementerror}
[\boldsymbol{r}_{D}]_i
=\left\Vert \boldsymbol{q}_i - \left(\boldsymbol{p} + \boldsymbol{v}  \Delta t_i\right)\right\Vert  -\left\Vert \boldsymbol{q}_i - \boldsymbol{p}\right\Vert 
\text{.}
\end{align}

\hypertarget{R2}{\textbf{Remark 2}}: The conventional LSPM-D is a biased estimator for a moving UD. The position bias of the conventional LSPM-D grows unlimited with increasing UD speed, and the derivation is similar to Remark 1.

We denote the RMSE of the LSPM-D by $RMSE_D$:
\begin{equation} \label{eq:RMSED}
RMSE_D =\sqrt{\Vert\boldsymbol{\mu}_D\Vert^2+\mathrm{tr}(\bm{Q}_D)} \text{,}
\end{equation}
where $\bm{Q}_D=\bm{Q}_K$.

We note that for the LSPM-KVD with true known velocity, the position bias is zero. Therefore, we have
\begin{align} \label{eq:errorcompare}
RMSE_K \leq RMSE_D \text{,}
\end{align}
in which, if and only if the UD velocity is zero, i.e., the UD is stationary, the two RMSEs are equal.

Based on the comparison above, we can see that the parameters to be estimated for both the LSPM-KVD and the LSPM-D are identical, and the localization performances of both methods when the UD velocity is zero are the same. However, when the UD is moving, the positioning accuracy of the proposed LSPM-KVD is better than that of the LSPM-D. Thus, the LSPM-D can be treated as a special case of the LSPM-KVD when the UD velocity is zero.

\subsection{LSPM-UVD Position Error Analysis}
For the LSPM-UVD, similar to the LSPM-KVD, the position bias is
\begin{equation}
\boldsymbol{\mu}_U=\boldsymbol{0} \text{.}
\end{equation}
The position variance is given by
\begin{equation}
\bm{Q}_U=\left[(\bm{G}_U^T\bm{W}_{\rho}\bm{G}_U)^{-1}\right]_{1:N,1:N} \text{.}
\end{equation}
The position RMSE is expressed as
\begin{equation}
RMSE_U =\sqrt{\Vert\boldsymbol{\mu}_U\Vert^2+\mathrm{tr}(\bm{Q}_U)} \text{.}
\end{equation}
The velocity estimation error can be obtained similarly. For example, the velocity error variance is the $(N+3):(2N+2),(N+3):(2N+2)$ sub-matrix of $(\bm{G}_U^T\bm{W}_{\rho}\bm{G}_U)^{-1}$. We aim to focus on the position error, and thus the velocity error is not further discussed in this paper.

We denote the CRLB and FIM of the LSPM-UVD by $\mathsf{CRLB}_U$ and $\mathcal{F}_U$, respectively. Similar to the derivation of the CRLB for the LSPM-UVD, we obtain the $\mathsf{CRLB}_U$ as given by
\begin{equation} \label{eq:UVDCRLBFisher}
\mathsf{CRLB}_U([\boldsymbol{\theta}_U]_i)=\left[\mathcal{F}_U^{-1}\right]_{i,i}=\left[\left(\bm{G}_U^T\bm{W}_{\rho}\bm{G}_U\right)^{-1} \right]_{i,i}\text{.}
\end{equation}

We partition the $\bm{G}_U$ matrix as
\begin{equation}\label{eq:GUpartition}
\bm{G}_U=\left[\bm{G}_0 , \bm{G}_1 \right] \text{,}
\end{equation}
where $	[\bm{G}_0]_{i,:}=[-\boldsymbol{e}_{Ui}^T,1,\Delta t_i] $, and $ [\bm{G}_1]_{i,:}=-\boldsymbol{e}_{Ui}^T\Delta t_i$.

Then, $\mathcal{F}_U$ in (\ref{eq:UVDCRLBFisher}) is rewritten as
\begin{equation} \label{eq:UVDFIMvsGpartition}
\mathcal{F}_U =
\begin{bmatrix}
\bm{G}_0^T\bm{W}_{\rho}\bm{G}_0 &  \bm{G}_0^T\bm{W}_{\rho}\bm{G}_1 \\
\bm{G}_1^T\bm{W}_{\rho}\bm{G}_0 &  \bm{G}_1^T\bm{W}_{\rho}\bm{G}_1
\end{bmatrix} \text{.}
\end{equation}

The top-left $(N+2)\times (N+2)$ sub-matrix of the inverse matrix of FIM relating to $\bm{G}_0$ contains the position-related terms, which is derived as 
\begin{equation} \label{eq:KVDFIMinvGpart}
\begin{split}
&[\mathcal{F}_U^{-1}]_{1:(N+2),1:(N+2)} \\
&=\left(\bm{G}_0^T\bm{W}_{\rho}\bm{G}_0-\bm{G}_0^{T}\bm{W}_{\rho}\bm{G}_1\left(\bm{G}_1^{T}\bm{W}_{\rho}\bm{G}_1\right)^{-1}\bm{G}_1^{T}\bm{W}_{\rho}\bm{G}_0\right)^{-1} \text{.}
\end{split}
\end{equation}

Based on (\ref{eq:KVDFIMinvGpart}), we have
\begin{equation} \label{eq:UVDinequal}
[\mathcal{F}_{U}^{-1}]_{1:(N+2),1:(N+2)} \succ \left(\bm{G}_0^T\bm{W}_{\rho}\bm{G}_0\right)^{-1}\text{,}
\end{equation}
which is proved in Appendix \ref{Appendix2}.

We note that the design matrix $\bm{G}_K$ of the LSPM-KVD equals to $\bm{G}_0$, and thus we have $\mathcal{F}_K=\bm{G}_0^T\bm{W}_{\rho}\bm{G}_0$.

Then, (\ref{eq:UVDinequal}) becomes
\begin{equation} \label{eq:FIMKVDvsUVD}
[\mathcal{F}_{U}^{-1}]_{1:(N+2),1:(N+2)} \succ \mathcal{F}_K^{-1}\text{,}
\end{equation}
which shows that the position-related CRLB of the LSPM-UVD is larger than that of the LSPM-KVD.

\subsection{LSPM-PVD Position Error Analysis}\label{VDPositionError}
The position bias of the LSPM-PVD is given by
\begin{equation}
\boldsymbol{\mu}_P=\boldsymbol{0}\text{,}
\end{equation}
with a Gaussian distributed UD velocity.

The position variance reads
\begin{equation}
\bm{Q}_P=\left[(\bm{G}_P^T\bm{W}_{\rho}\bm{G}_P)^{-1}\right]_{1:N,1:N} \text{.}
\end{equation}
The RMSE is 
\begin{equation}
RMSE_P =\sqrt{\Vert\boldsymbol{\mu}_P\Vert^2+\mathrm{tr}(\bm{Q}_P)} \text{.}
\end{equation}

We now derive the CRLB of the LSPM-PVD. After some similar derivation to that in Section \ref{ErrorKVD}, the FIM of the LSPM-PVD, denoted by $\mathcal{F}_P$, is 
\begin{equation}\label{eq:FIMvsG}
\mathcal{F}_P = \bm{G}_P^T\bm{W}\bm{G}_P \text{,}
\end{equation}
where $\bm{G}_P$ is given by (\ref{eq:Gderivative}) and $\bm{W}$ is given by (\ref{eq:weightmat}).

The CRLB for the LSPM-PVD is given by
\begin{equation} \label{eq:CRLBFisher}
\mathsf{CRLB}_P([\boldsymbol{\theta}]_i)=[\mathcal{F}_P^{-1}]_{i,i} \text{,}
\end{equation}



Furthermore, we investigate the relationship between the errors of the LSPM-PVD, LSPM-KVD and LSPM-UVD. We partition the $\bm{G}_P$ matrix as given by
\begin{equation}\label{eq:Gpartition}
\bm{G}_P=\left[
\begin{matrix}
\bm{G}_0 & \bm{G}_1 \\
\bm{O}_{N\times (N+2)} & \bm{I}_{N}
\end{matrix}
\right] \text{,}
\end{equation}
where $\bm{G}_0$ and $\bm{G}_1$ have the same definition as (\ref{eq:GUpartition}).

We define the velocity-related weighting matrix as $\bm{W}_{v} \triangleq \bm{\Sigma}_{v}^{-1}$. Then, the FIM in (\ref{eq:FIMvsG}) is derived as
\begin{equation} \label{eq:FIMvsGpartition}
\mathcal{F}_P =
\begin{bmatrix}
\bm{G}_0^T\bm{W}_{\rho}\bm{G}_0 &  \bm{G}_0^T\bm{W}_{\rho}\bm{G}_1 \\
\bm{G}_1^T\bm{W}_{\rho}\bm{G}_0 &  \bm{G}_1^T\bm{W}_{\rho}\bm{G}_1+\bm{W}_v
\end{bmatrix} \text{.}
\end{equation}

The $\bm{G}_0$ related sub-matrix in the inverse of the FIM is derived as 
\begin{equation} \label{eq:FIMinvGpart}
\begin{split}
[\mathcal{F}_P^{-1}]_{1:(N+2),1:(N+2)} =\left(\bm{G}_0^T\bm{W}_{\rho}\bm{G}_0-\bm{D}\right)^{-1} \text{,}
\end{split}
\end{equation}
where $\bm{D}=\bm{G}_0^{T}\bm{W}_{\rho}\bm{G}_1\left(\bm{G}_1^{T}\bm{W}_{\rho}\bm{G}_1+\bm{W}_v\right)^{-1}\bm{G}_1^{T}\bm{W}_{\rho}\bm{G}_0$.

Based on similar proof to Appendix \ref{Appendix2}, we come to
\begin{eqnarray}
\mathcal{F}_K^{-1}\prec [\mathcal{F}_P^{-1}]_{1:(N+2),1:(N+2)}\prec[\mathcal{F}_U^{-1}]_{1:(N+2),1:(N+2)} \text{,}
\end{eqnarray}
which shows that the position related CRLB of the proposed LSPM-PVD lies between that of the LSPM-KVD and LSPM-UVD.

\section{Numerical Simulation} \label{simulation}
Simulations are conducted in this section to evaluate the localization performance of the proposed LSPM-KVD, LSPM-UVD and LSPM-PVD. In all the simulations, we compute the RMSE of the positioning results. The CRLB is used as a metric to evaluate the positioning accuracy. The RMSE of the positioning results is given by
\begin{align} \label{eq:RMSEdef}
RMSE&=\sqrt{\frac{1}{N_s}\sum_{1}^{N_s}\Vert\boldsymbol{p}-\hat{\boldsymbol{p}}\Vert^2}
\end{align}
where $N_s$ is the total number of positioning result samples and $\hat{\boldsymbol{p}}$ is the localization result from the algorithm under test for each simulated sample.

\subsection{LSPM-KVD Performance Evaluation: Stationary UD} \label{2Dstationary}
\begin{figure}
	\centering
	\includegraphics[width=0.99\linewidth]{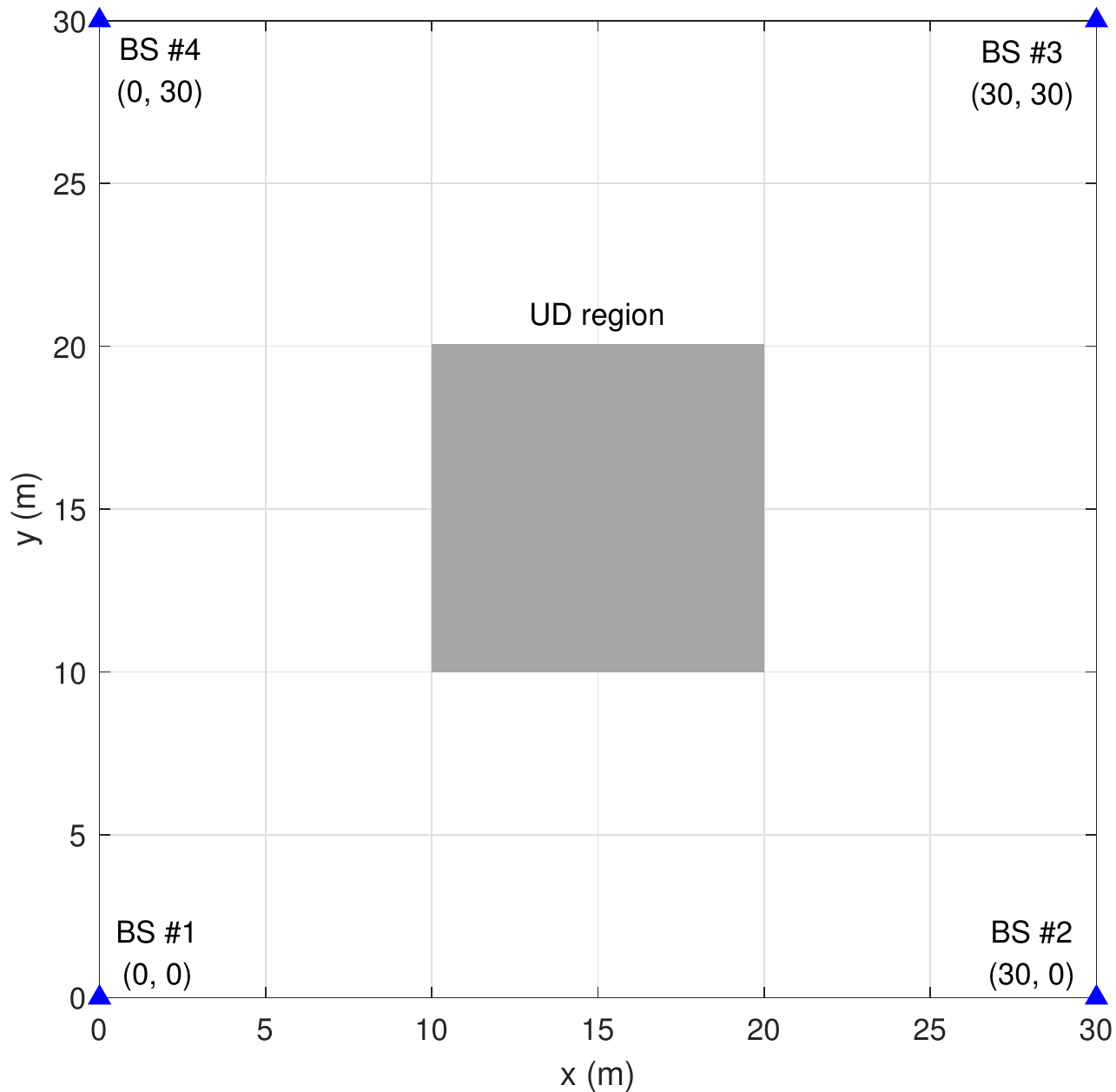}
	\caption{BSs and UD positions for simulation. BSs are located at the corner and UD is randomly placed in the gray square region. }
	\label{fig:simulationsetting}
\end{figure}

A simulation scenario in 2D is first created to evaluate the performance of the LSPM-KVD, LSPM-UVD and LSPM-PVD in different cases. Four BSs are placed at the corners of a 30 m$\times$30 m square area as shown in Fig. \ref{fig:simulationsetting}. One UD receives signals and obtains pseudorange measurements from BSs in a sequential manner with a constant interval of 0.01 s between two consecutive measurements. The UD is stationary or moving inside the square area with a side length of 10 m and center at (15, 15) m, as shown in Fig. \ref{fig:simulationsetting}. The relative clock drift between the UD and BSs is set to 5 parts per million (ppm), which is at the level of a smartphone-used temperature compensated crystal oscillator (TCXO) \cite{lam2016review}. The number of input pseudorange measurements to the algorithms is set to be $M=8$, i.e., the UD performs one position fix after having obtained two rounds of pseudorange measurements from the four BSs sequentially in this simulation. The localization time instant is set as the first measurement time, i.e., $t_L=t_1$. The maximum iteration time $iter$ is set to be 20 and the convergence threshold $thr$ is set to be 10$^{-3}$.

\begin{figure}
	\centering
	\includegraphics[width=0.99\linewidth]{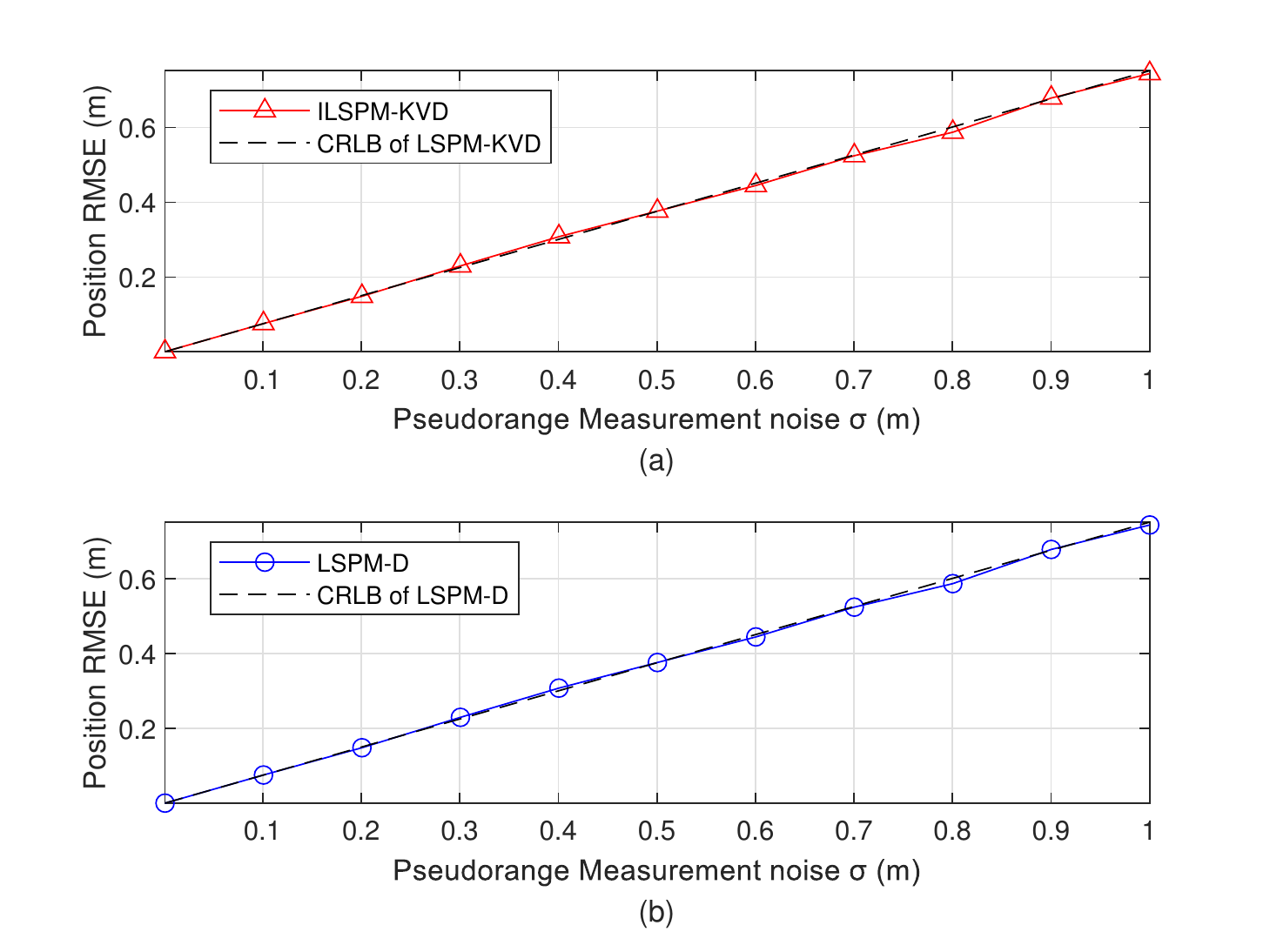} 
	\caption{Position error vs. measurement noise in stationary case. The positioning accuracy of the proposed ILSPM-KVD algorithm reaches CRLB, and is identical with that of the LSPM-D, which also reaches its CRLB for a stationary UD.
	}
	\label{fig:staticresult}
\end{figure}

A stationary simulation scene is created to evaluate the performance of the proposed LSPM-KVD method. The UD is randomly placed inside the gray square area as shown in Fig. \ref{fig:simulationsetting}. All the TOA measurement noise $\sigma_i$ is set identical, varying from 0.01 to 1 m. We run 1000 times of Monte Carlo simulations at every step to generate the UD position and the sequential measurements. The localization error result of the proposed ILSPM-KVD algorithm is shown in Fig. \ref{fig:staticresult} (a). The CRLB is shown in the same figure. It can be observed from the figure that the theoretical error is identical with the numerical position RMSE result given by the ILSPM-KVD algorithm. This also indicates that in such a stationary case, the proposed algorithm is an unbiased estimator.

We also use the conventional LSPM-D method to compute the stationary positioning results, which is depicted in Fig. \ref{fig:staticresult} (b). The CRLB of the LSPM-D method is also computed and plotted in the same figure. We can see that the CRLB of the LSPM-D is identical with that of the LSPM-KVD. This result shows that the proposed LSPM-KVD method provides the same positioning accuracy as the conventional LSPM-D method in the stationary case. Thus, the conventional LSPM-D method can be considered as a special case in the proposed LSPM-KVD method. It conforms to the error analysis in Section \ref{comparison}.

\subsection{LSPM-KVD Performance Evaluation: Moving UD}
We conduct a simulation to evaluate the performance of the LSPM-KVD for a moving UD. All the simulation settings are identical with that of Section \ref{2Dstationary}, except that the UD randomly placed in the same square area shown in Fig. \ref{fig:simulationsetting} is moving with a constant velocity during one period of localization with $M$ measurements. We set the true speed of the UD as 5 m/s with a randomly chosen direction during the entire measurement period. The TOA measurement noise $\sigma$ is set varying from 0.01 to 1 m. At every noise step, 1000 simulation runs are done.

\begin{figure}
	\centering
	\includegraphics[width=0.99\linewidth]{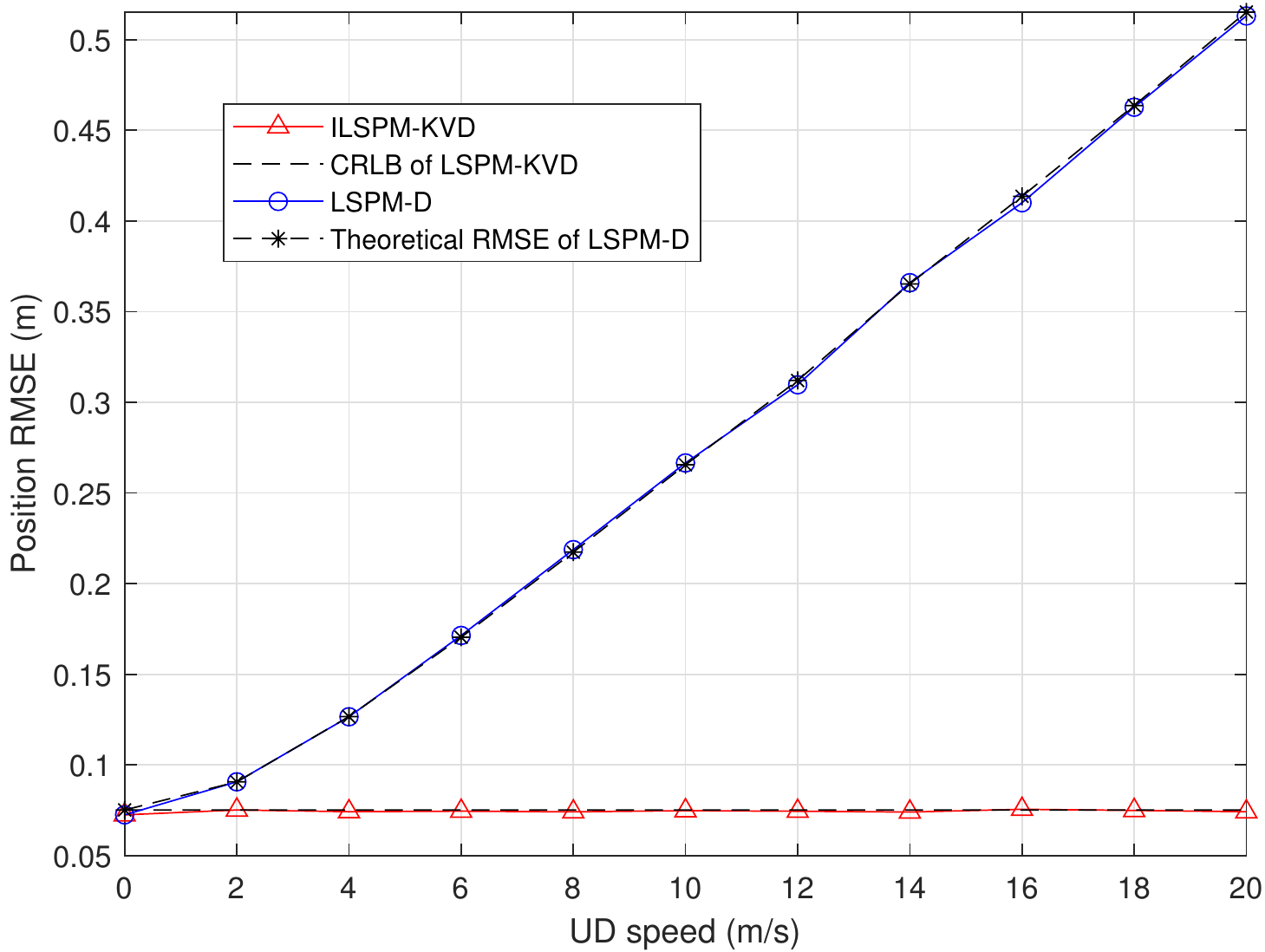} 
	\caption{Position error vs. UD speed ($\sigma$=0.1 m). The position error of the proposed ILSPM-KVD remains stable with increasing UD speed and reaches CRLB. The conventional LSPM-D method has an increasing position bias when the UD speed increases. The position error of the LSPM-D matches the theoretical analysis.
	}
	\label{fig:errorvsdP1}
\end{figure}

We conduct simulations with different user speeds to investigate the localization performance of the proposed ILSPM-KVD. We set the measurement noise to $\sigma$=0.1 m, which is at the level of the measurement capability of a UWB device \cite{ruiz2017comparing}, and vary the user speed from 0.1 m/s to 20 m/s. The position error result versus the UD speed is illustrated in Fig. \ref{fig:errorvsdP1}. It can be observed that when the UD speed increases, the proposed ILSPM-KVD has stable localization error, which matches the CRLB and shows that it is an unbiased estimator. On the contrary, the localization error of the LSPM-D method grows with an increasing user speed. When the UD speed is small, the noise-caused position error dominants the total error of the LSPM-D. With larger speed, the localization error increases approximately linearly due to the dominance of the velocity-caused position bias. This result is consistent with \hyperlink{R2}{Remark 2}. The position RMSE from the LSPM-D matches the theoretical RMSE computed based on (\ref{eq:RMSED}) as depicted in the same figure, verifying the error analysis for the conventional LSPM-D in Section \ref{comparison}.

\begin{figure}
	\centering
	\includegraphics[width=0.99\linewidth]{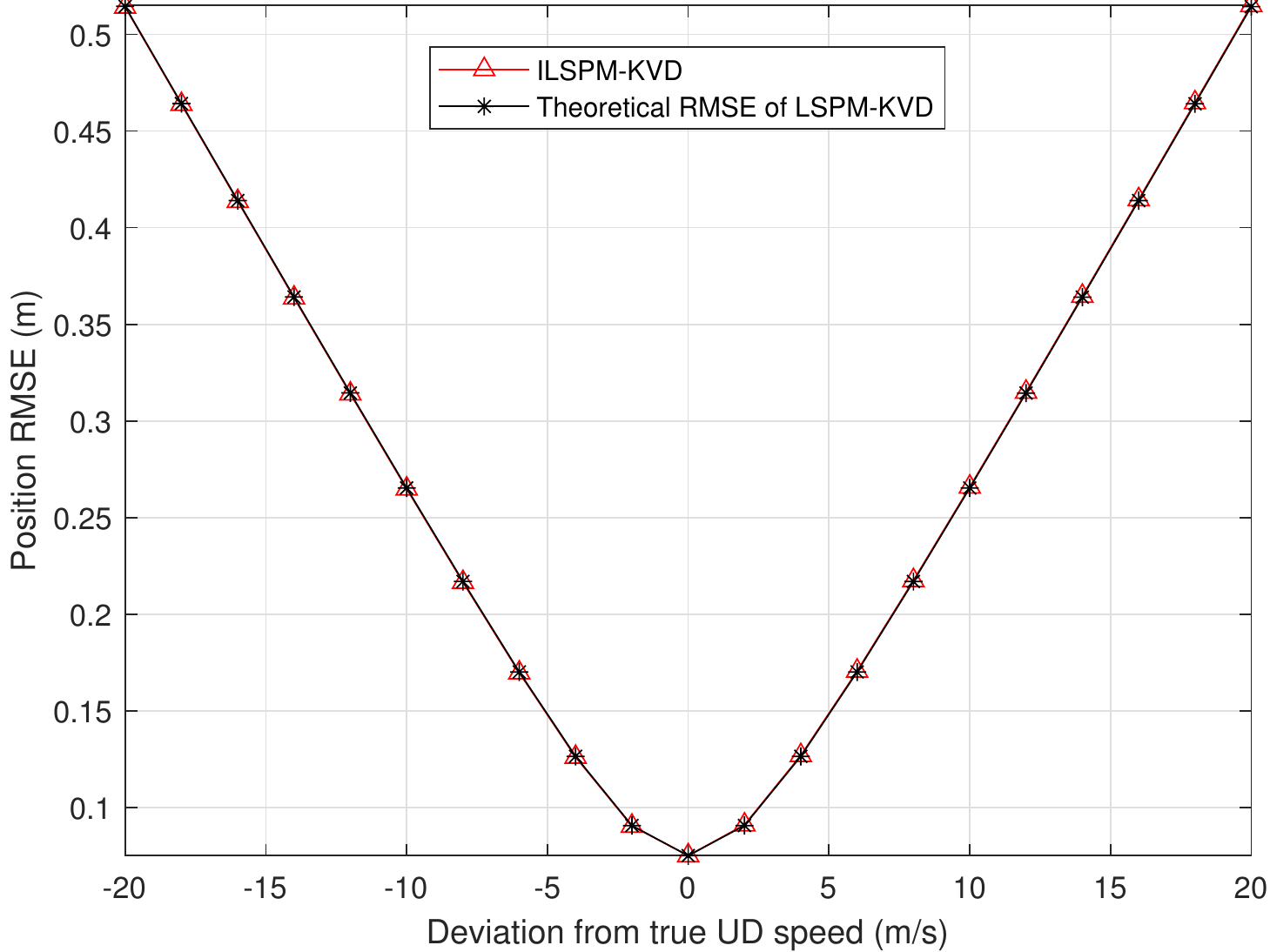} 
	\caption{Position error vs. deviation from true UD speed ($\sigma$=0.1 m and true UD speed is 5 m/s). The position error of the proposed ILSPM-KVD increases when the assumed known velocity deviates from the true velocity. The RMSE output by the ILSPM-KVD matches the theoretical analysis.
	}
	\label{fig:errorvsdV}
\end{figure}

When the assumed known velocity is not accurate enough and deviates from the true UD velocity, the localization error of the LSPM-KVD increases as shown in Fig. \ref{fig:errorvsdV}. We set the measurement noise as $\sigma$=0.1 m, and the true UD speed as 5 m/s.  The position RMSE increases when the assumed UD speed deviates from its true value. When the UD speed becomes larger, the growing of the position RMSE appears more linear. The theoretical RMSE at each simulation run is computed using (\ref{eq:RMSEdV}) and is found to match the position RMSE curve well, verifying the error analysis in Section \ref{ErrorKVD}.

\subsection{Performance Evaluation of LSPM-UVD and LSPM-PVD} \label{UVDVD}
Similar to the moving simulation for the LSPM-KVD, the UD position is randomly selected in the UD region shown in Fig. \ref{fig:simulationsetting}. We set the true speed of the UD as 5 m/s with a randomly chosen direction during one localization period. The TOA measurement noise $\sigma$ is also set varying from 0.01 to 1 m. At every noise step, 1000 simulation runs are done. As for the LSPM-PVD evaluation, we set that the UD velocity follows a Gaussian distribution with a standard deviation (STD) of 2 m/s for each axis in the simulation.

\begin{figure}
	\centering
	\includegraphics[width=0.99\linewidth]{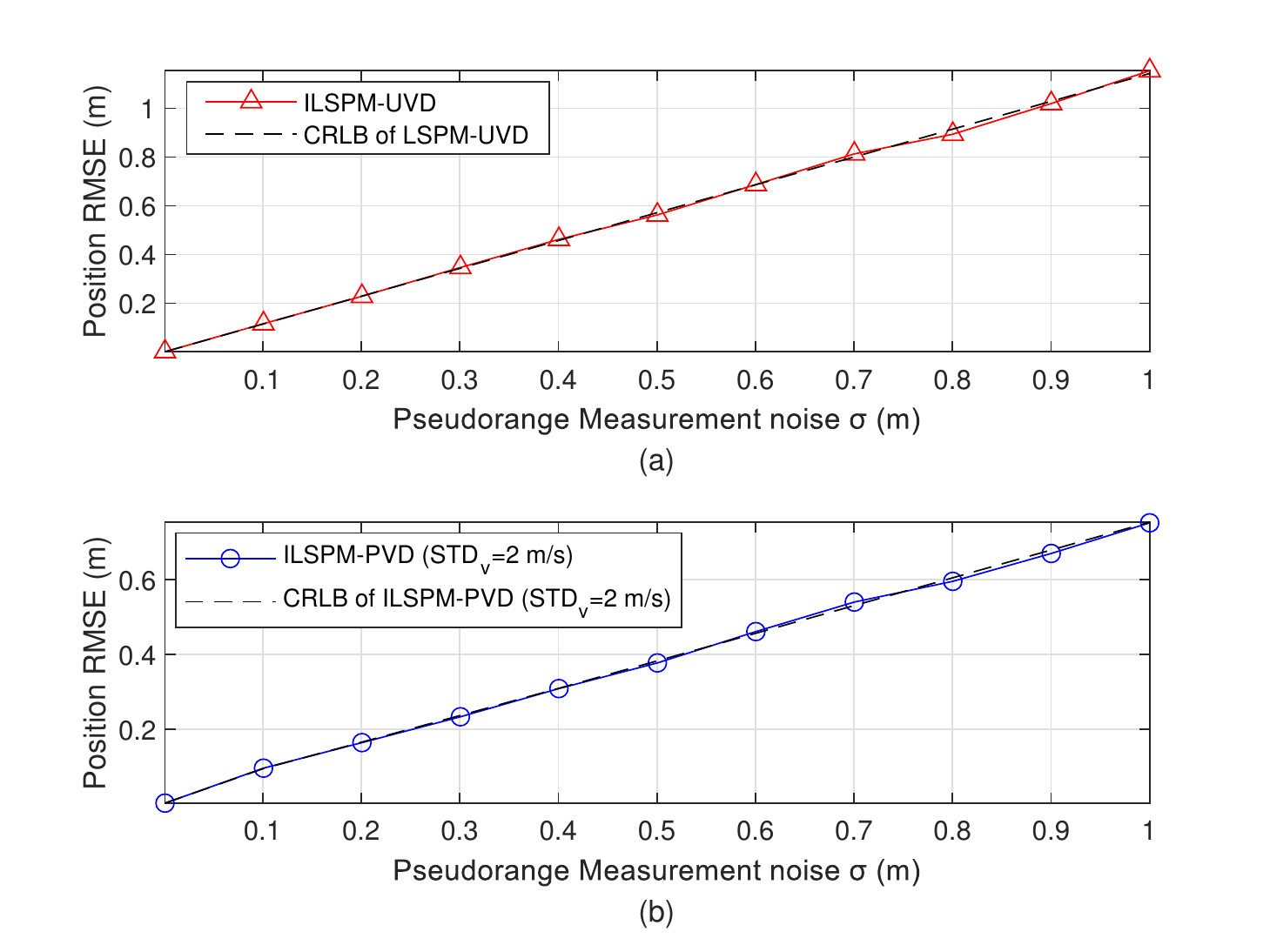} 
	\caption{Position error vs. pseudorange measurement noise for the LSPM-UVD (a) and LSPM-PVD (b). The true UD speed is set to 5 m/s. For the LSPM-PVD, the mean velocity is set to 5 m/s, and the STD is set to 2 m/s. Positioning accuracy of both the positioning algorithms reaches CRLB.}
	\label{fig:resultUandVD}
\end{figure}

The position error results of both methods are shown in Fig \ref{fig:resultUandVD} (a) and (b), respectively. Their CRLBs are depicted in the same figure. We can see that the localization accuracy of both the LSPM-UVD and LSPM-PVD methods reach their CRLB, showing that they are unbiased estimators.

\begin{figure}
	\centering
	\includegraphics[width=0.99\linewidth]{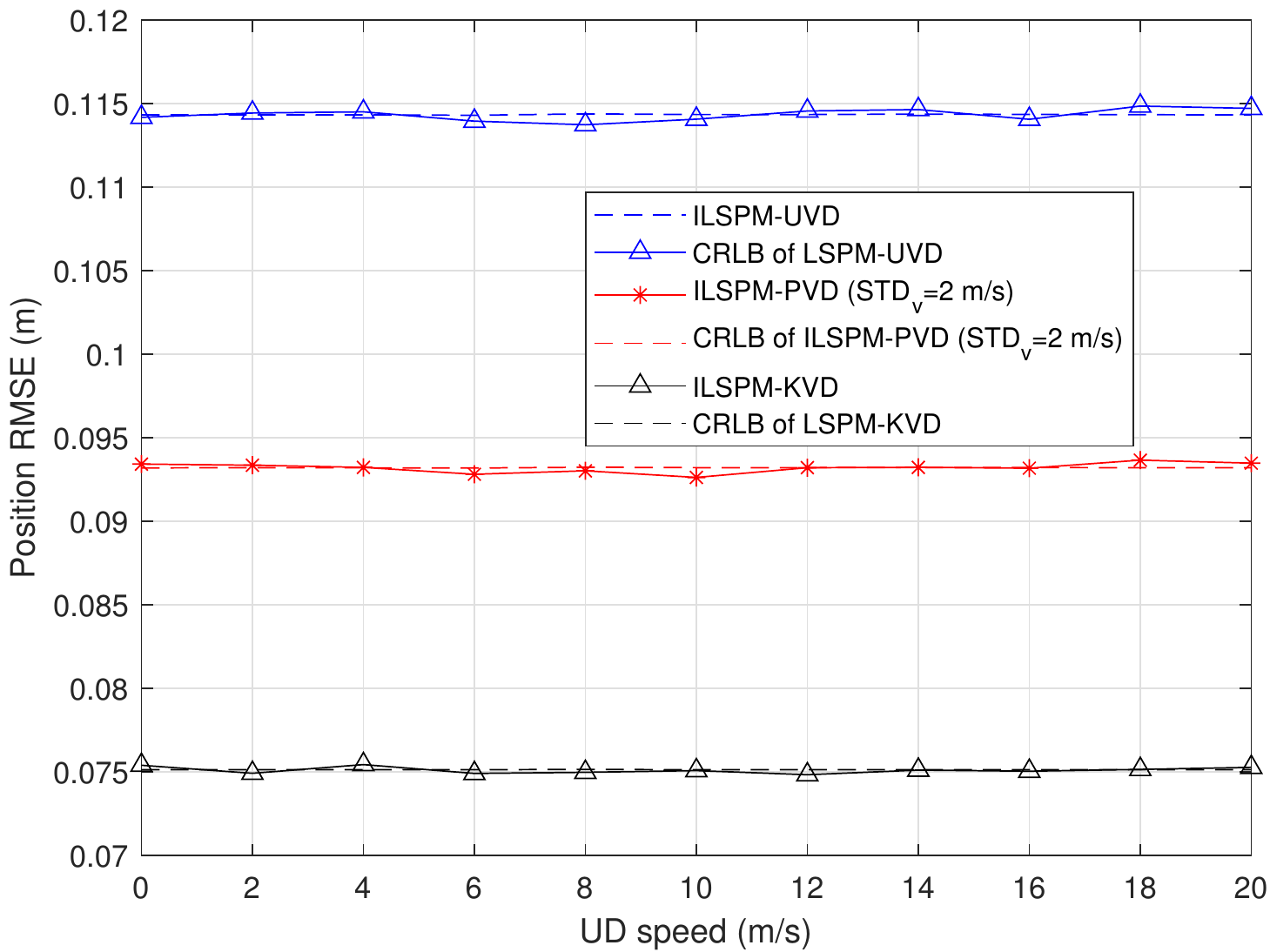} 
	\caption{Position error vs. UD velocity. Measurement noise $\sigma$=0.1 m. For the LSPM-PVD, the prior distribution of the UD velocity is set to have a mean at the true velocity and a STD of 2 m/s. All the three proposed methods reaches CRLB. Positioning accuracy of the LSPM-PVD lies between that of the LSPM-KVD and LSPM-UVD.}
	\label{fig:comparisonUKP}
\end{figure}

In order to evaluate the performance with varying velocity, we then fix the measurement noise to $\sigma$=0.1 m, and vary the UD speed from 0 to 20 m/s. The prior velocity distribution input to the LSPM-PVD method has a mean value at the true velocity and a STD of 2 m/s. The position error result is shown in Fig. \ref{fig:comparisonUKP}. We also include the result of the LSPM-KVD in the same figure. We can see that with prior knowledge on the UD velocity, the performance of the LSPM-PVD is better than that of the LSPM-UVD. The prior knowledge of the velocity for the LSPM-PVD is not as accurate as the true velocity for the LSPM-KVD, and thus the position error of the LSPM-PVD is larger than that of the LSPM-KVD, as expected. The LSPM-UVD does not require any knowledge of the UD velocity, indicating a better robustness.

\subsection{Localization Performance in Circular Motion Scenario} \label{circularmotion}
\begin{figure}
	\centering
	\includegraphics[width=0.99\linewidth]{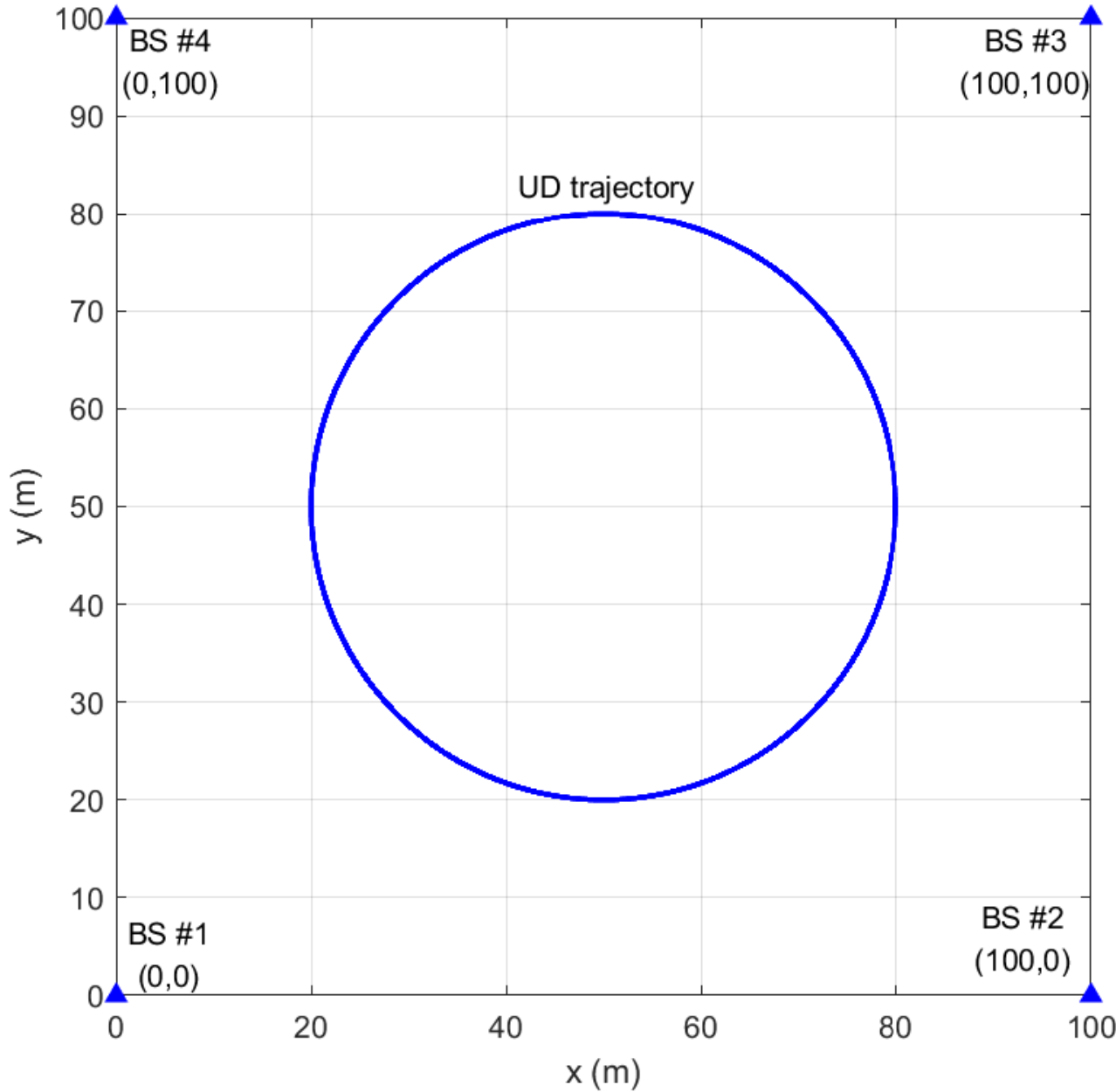} 
	\caption{BS placement and UD trajectory for circular motion case.}
	\label{fig:settingcircle}
\end{figure}

We investigate the localization performance of the proposed methods in another simulation scenario that the UD conducts a circular motion. This scenario is often seen in quadrotor control and flight. There are four BSs placed at the corners of a square area with a side length of 100 m as shown in Fig. \ref{fig:settingcircle}. We set the linear speed of the UD to 10 m/s, which can be achieved by a main-stream commercial quadrotor. The trajectory of the UD is a circle centered at (50, 50) m with a radius of 30 m as shown in Fig. \ref{fig:settingcircle}. The total simulation time length is 360 s. Other settings are identical with the previous simulation, i.e., the interval of sequential pseudorange measurements is set to 0.01 s, the number of input measurements is $M=8$, The relative clock drift between the UD and BSs is set to 5 ppm, and the pseudorange measurement noise is set to the level of a UWB device as $\sigma=0.1$ m. The simulated pseudorange measurements are input to all the three proposed methods as well as the conventional LSPM-D. For the LSPM-PVD, we set that the prior distribution of the UD velocity follows a Gaussian distribution with the mean of the true UD velocity and a STD of 2 m/s.

\begin{figure}
	\centering
	\includegraphics[width=0.99\linewidth]{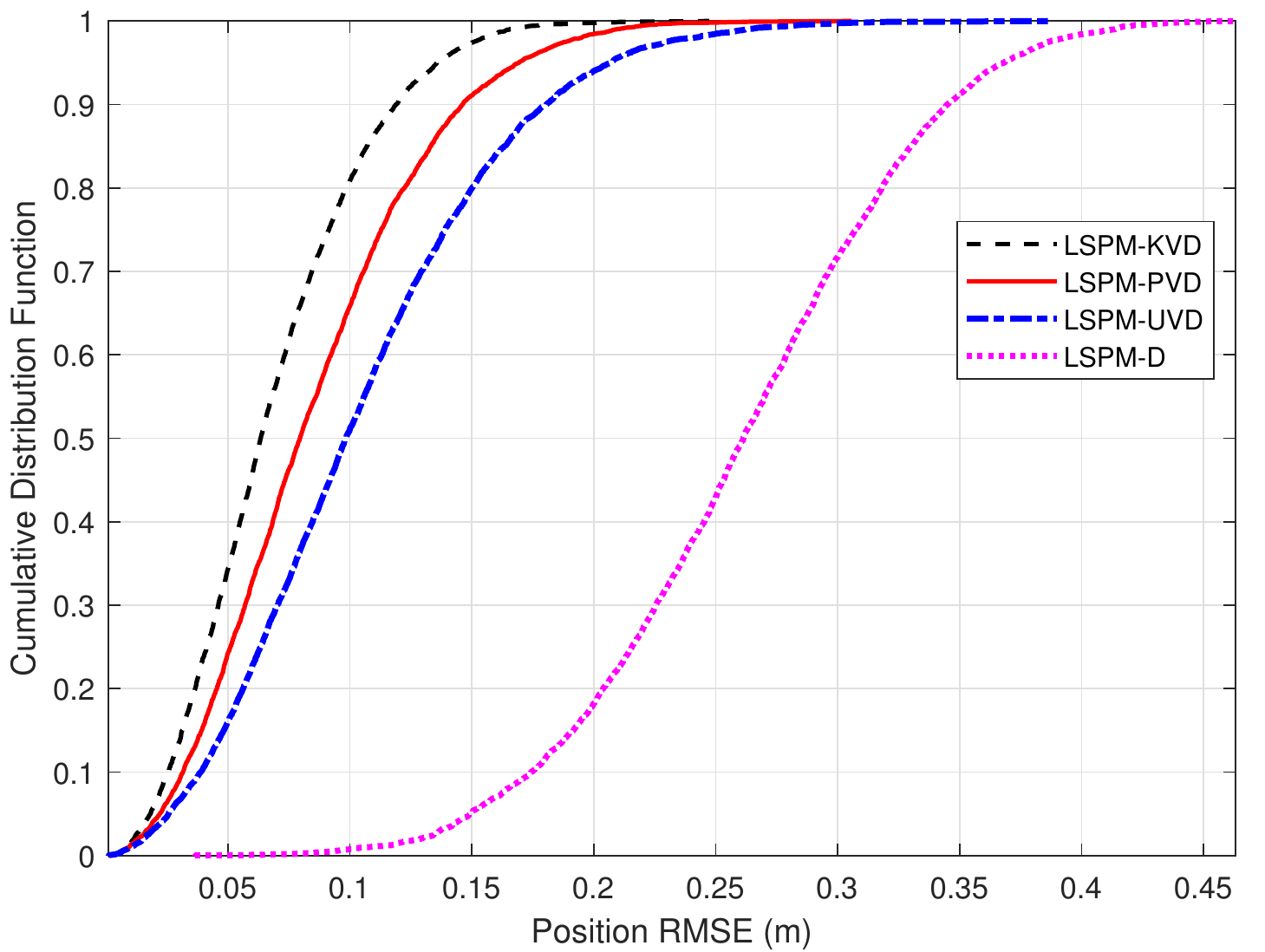} 
	\caption{CDF of the position RMSE in the simulated circular motion case. The true UD speed is set to 10 m/s. For the LSPM-PVD, the prior UD velocity is set to have a mean value at the true velocity and a STD of 2 m/s. The position RMSE of the LSPM-KVD is the smallest, the LSPM-PVD places second and the LSPM-UVD is the third best. The conventional LSPM-D method has the largest localization error due to the movement-caused bias.}
	\label{fig:totaldP}
\end{figure}

The cumulative distribution function (CDF) curves of the position error of all the three proposed methods and the conventional LSPM-D are plotted in Fig. \ref{fig:totaldP}. The $x$ and $y$-axis and the total position RMSEs are listed in Table \ref{table_circleresult}. We can see that the LSPM-KVD has the best localization accuracy among all the methods, and the LSPM-PVD and LSPM-UVD place second and third. The LSPM-D method has the worst localization accuracy due to the unmodeled UD motion. This result verifies the performance analysis given in Section \ref{locanalysis} and demonstrates the feasibility of the proposed methods in a real-world application such as drone flight tracking.

In terms of computational complexity, we compare the three proposed methods with the conventional LSPM-D. The LSPM-KVD has the same number of estimated parameters as the LSPM-D, and thus their computational loads are identical. The LSPM-UVD and LSPM-PVD have extra UD velocity to be estimated, and thus their computational complexity is larger than the LSPM-D. The most computation-intensive step of the LSPM-UVD and LSPM-PVD is the $(2N+2)\times(2N+2)$ matrix inverse compared with the $(N+2)\times(N+2)$ ($N$=3 in a 3D localization case) matrix inverse of the LSPM-D. For a commercially available embedded system, this task of at most 8$\times$8 matrix inverse can be completed in a real-time manner with acceptable computational resources \cite{cao2018indoor}, indicating the feasibility of the proposed methods in light-weighted equipment such as wearable sensors and Internet of Things (IoT) devices.

\begin{table}[!t]
	\begin{threeparttable}
	\caption{Localization error result of circular motion simulation}
	\label{table_circleresult}
	\centering
	\begin{tabular}{c c c c c }
		\toprule
		RMSE & LSPM-KVD & LSPM-PVD & LSPM-UVD & LSPM-D\\
		\hline
		$x$-axis (cm) & 5.2 & 6.5& 8.0& 19.0\\
		$y$-axis (cm) & 5.9 & 7.2& 8.9& 19.1\\
		position (cm) & 7.8 & 9.7& 12.0& 26.9\\
		\bottomrule
	\end{tabular}

\begin{tablenotes}[para,flushleft]
Note: The UD has a constant speed of 10 m/s. For the LSPM-PVD, the prior UD velocity is set to have a mean at the true velocity and its STD is set to 2 m/s. The LSPM-KVD with perfectly known velocity has the best positioning accuracy. The position error of the LSPM-PVD is larger than the LSPM-KVD but smaller than the LSPM-UVD. The LSPM-UVD has the largest position error among the three proposed methods since it uses least prior knowledge on the velocity. The conventional LSPM-D has the worst positioning accuracy due to the movement-caused bias.	
\end{tablenotes}
\end{threeparttable}
\end{table}

\section{Conclusion}
In the TDBPS, the BSs transmit signals successively and pseudorange measurements are obtained sequentially by a UD. When the conventional LSPM-D method is directly used, the position error grows with an increasing UD speed due to lack of estimation and compensation of the UD velocity. To solve this movement-caused error problem, in this paper, we present a set of optimal localization methods utilizing different levels of prior knowledge on the UD velocity. We first propose the LSPM-KVD method to incorporate the known UD velocity to reduce the displacement error and show that the conventional LSPM-D is a special case of the LSPM-KVD when the UD is stationary. To deal with the case of unknown UD velocity, the LSPM-UVD method that jointly estimates the UD position and velocity is proposed. For a more general case when the prior distribution of the UD velocity is available, we develop the LSPM-PVD method, a MAP estimator and a generalized form of the LSPM-KVD and LSPM-UVD. We conduct localization performance analysis of the three proposed methods. We derive their CRLBs and position errors. Performance analysis shows that i) the LSPM-KVD outperforms the conventional LSPM-D in terms of localization accuracy, ii) the position error of the LSPM-KVD increases when the assumed known UD velocity deviates from the true value, and iii) the LSPM-PVD method has better robustness and larger position error than the LSPM-KVD, and the LSPM-UVD does not require any prior knowledge on the UD velocity and thus has the best robustness among the three. Simulations on stationary and moving cases verify the theoretical analysis. Results show that with accurately known UD velocity, the localization accuracy of the LSPM-KVD is optimal. With a prior distribution on the UD velocity, the LSPM-PVD is optimal and has larger position error than the LSPM-KVD. In the case of unknown UD velocity, the LSPM-UVD is the optimal estimator. It does not require any prior knowledge on the UD velocity, leading to best robustness. The three proposed localization methods are able to be implemented in light-weighted real-time systems and are feasible for real applications in the TDBPS.

\appendices
\section{Derivation of Remark 1}
\label{Appendix1}
In the case with deviated known UD velocity, according to (\ref{eq:dPvsdV}), the squared norm of the position bias is given by
\begin{equation}\label{eq:squaredbias}
\Vert\tilde{\boldsymbol{\mu}}_{K}\Vert^2=\tilde{\boldsymbol{r}}_K^T\bm{S}_1^T\bm{S}_1\tilde{\boldsymbol{r}}_K \text{,}
\end{equation}
where $\bm{S}_1=\left[(\tilde{\bm{G}}_K^T\bm{W}_{\rho}\tilde{\bm{G}}_K)^{-1}\tilde{\bm{G}}_K^T\bm{W}_{\rho}\right]_{1:N,:}$.

We conduct Taylor series expansion on $\boldsymbol{r}_{K}$ given by (\ref{eq:resdV}) at the assumed velocity $\tilde{\boldsymbol{v}}$ and keep the first order term, and come to
\begin{align} \label{eq:rexpansion}
\tilde{\boldsymbol{r}}_{K}=\bm{S}_2\Delta\boldsymbol{v} \text{,}
\end{align}
where
$
[\bm{S}_2]_{i,:}=\frac{\Delta t_i\left(\boldsymbol{q}_i - \boldsymbol{p} \right)^T}{\left\Vert \boldsymbol{q}_i - \boldsymbol{p}\right\Vert}
$.

By substituting (\ref{eq:rexpansion}) into (\ref{eq:squaredbias}), we have
\begin{align}\label{eq:squarebias1}
\Vert\tilde{\boldsymbol{\mu}}_{K}\Vert^2=\Delta\boldsymbol{v}^T\bm{S}\Delta\boldsymbol{v}\text{,}
\end{align}
where $\bm{S}=\bm{S}_2^T\bm{S}_1^T\bm{S}_1\bm{S}_2$.

We note that $\bm{S}$ is positive definite. We can thereby find a positive scalar $\alpha$ to let the matrix $\bm{S}-\alpha\bm{I}$ be positive semi-definite. Therefore, we come to
\begin{align}\label{eq:squareerror2}
\Vert\tilde{\boldsymbol{\mu}}_{K}\Vert^2= \alpha\Delta\boldsymbol{v}^T\Delta\boldsymbol{v} + \Delta\boldsymbol{v}^T\left(\bm{S}-\alpha\bm{I}\right)\Delta\boldsymbol{v} \geq\alpha\Vert \Delta\boldsymbol{v}\Vert^2
\end{align}

We can see from (\ref{eq:squareerror2}) that the position bias grows approximately linearly with an increasing UD speed deviation.

\section{Proof of Inequality (\ref{eq:UVDinequal})}
\label{Appendix2}

We rewrite (\ref{eq:KVDFIMinvGpart}) here as
\begin{equation} \label{eq:submatG0}
\begin{split}
&[\mathcal{F}_U^{-1}(\boldsymbol{\theta})]_{1:(N+2),1:(N+2)} \\
&=\left(\bm{G}_0^T\bm{W}_{\rho}\bm{G}_0-\bm{G}_0^{T}\bm{W}_{\rho}\bm{G}_1\left(\bm{G}_1^{T}\bm{W}_{\rho}\bm{G}_1\right)^{-1}\bm{G}_1^{T}\bm{W}_{\rho}\bm{G}_0\right)^{-1} \text{.}
\end{split}
\end{equation}

We know that $\bm{W}_{\rho}$ is a diagonal matrix with all diagonal entries being positive values. In practice, the matrices $\bm{G}_0$ and $\bm{G}_1$ usually have full rank if there are sufficient number of observed BSs. Because $\bm{A}\bm{A}^T$ is a positive definite matrix for an arbitrary real matrix $\bm{A}$ with full rank, both $\bm{G}_0^T\bm{W}_{\rho}\bm{G}_0$ and $\bm{G}_1^T\bm{W}_{\rho}\bm{G}_1$ are positive definite. Furthermore, $\bm{G}_0^{T}\bm{W}_{\rho}\bm{G}_1\left(\bm{G}_1^{T}\bm{W}_{\rho}\bm{G}_1\right)^{-1}\bm{G}_1^{T}\bm{W}_{\rho}\bm{G}_0$ is also positive definite.

Note that $\bm{A}\succ\bm{B}$ if and only if $(\bm{A}-\bm{B})$ is positive definite. Thus,
\begin{align} \label{eq:matgreater}
&\bm{G}_0^T\bm{W}_{\rho}\bm{G}_0\succ \nonumber\\
&\bm{G}_0^T\bm{W}_{\rho}\bm{G}_0-\bm{G}_0^{T}\bm{W}_{\rho}\bm{G}_1\left(\bm{G}_1^{T}\bm{W}_{\rho}\bm{G}_1\right)^{-1}\bm{G}_1^{T}\bm{W}_{\rho}\bm{G}_0 \text{.}
\end{align}

According to \cite{horn2012matrix}, we apply inverse operation to the matrices on both sides of (\ref{eq:matgreater}) and come to
\begin{align} \label{eq:matsmaller}
&\left(\bm{G}_0^T\bm{W}_{\rho}\bm{G}_0\right)^{-1}\prec \nonumber\\
&\left(\bm{G}_0^T\bm{W}_{\rho}\bm{G}_0-\bm{G}_0^{T}\bm{W}_{\rho}\bm{G}_1\left(\bm{G}_1^{T}\bm{W}_{\rho}\bm{G}_1\right)^{-1}\bm{G}_1^{T}\bm{W}_{\rho}\bm{G}_0\right)^{-1} \text{.}
\end{align}

Finally, (\ref{eq:UVDinequal}) is proved.


%


\ifCLASSOPTIONcaptionsoff
    \newpage
\fi



\bibliographystyle{IEEEtran}
\bibliography{IEEEabrv,paper}
%


%









\end{document}